\documentclass[lettersize,journal]{IEEEtran}
\usepackage{amsmath,amsfonts}
\usepackage{amsmath,amssymb,bm}
\usepackage{mathrsfs}
\usepackage{algorithmic}
\usepackage{algorithm}
\usepackage{array}
\usepackage[caption=false,font=normalsize,labelfont=sf,textfont=sf]{subfig}
\usepackage{textcomp}
\usepackage{stfloats}
\usepackage{url}
\usepackage{verbatim}
\usepackage{graphicx}
\usepackage{cite}
\usepackage{multirow} 
\usepackage{booktabs}
\begin{document}

\title{Blind Quality Enhancement for G-PCC Compressed \\ Dynamic Point Clouds}

\author{Tian Guo, Hui Yuan,~\IEEEmembership{Senior Member,~IEEE,} Chang Sun, Wei Zhang,~\IEEEmembership{Senior Member,~IEEE}, Raouf Hamzaoui,~\IEEEmembership{Senior Member,~IEEE}, and Sam Kwong,~\IEEEmembership{Fellow,~IEEE}
\thanks{This work was supported in part by the National Natural Science Foundation of China under Grants 62571303, the High-end Foreign Experts Recruitment Plan of Chinese Ministry of Human Resources and Social Security under Grant H20251083. \textit{(Corresponding author: Hui Yuan)}} 
\thanks{Tian Guo, Hui Yuan, Chang Sun, and Wei Zhang are with the School of Control Science and Engineering, Shandong University, Ji'nan, 250061, China, and also with the Key Laboratory of Machine Intelligence and System Control, Ministry of Education, Ji'nan, 250061, China (e-mail: guotiansdu@mail.sdu.edu.cn; huiyuan@sdu.edu.cn; 202220795@mail.sdu.edu.cn; davidzhang@sdu.edu.cn).}
\thanks{Raouf Hamzaoui is with the School of Engineering, Infrastructure and Sustainability, De Montfort University, LE1 9BH Leicester, UK. (e-mail: rhamzaoui@dmu.ac.uk).}
\thanks{Sam Kwong is with the Department of Computing and Decision Science,
 Lingnan University, Hong Kong (e-mail: samkwong@ln.edu.hk).}
}

\markboth{Journal of \LaTeX\ Class Files,~Vol.~14, No.~8, August~2021}%
{Guo \MakeLowercase{\textit{et al.}}:STQE: Spatial-Temporal Attribute Quality Enhancement for G-PCC Compressed Dynamic Point Clouds}


\maketitle

\begin{abstract}
Point cloud compression often introduces noticeable reconstruction artifacts, which makes quality enhancement necessary. Existing approaches typically assume prior knowledge of the distortion level and train multiple models with identical architectures, each designed for a specific distortion setting. This significantly limits their practical applicability in scenarios where the distortion level is unknown and computational resources are limited. To overcome these limitations, we propose the first blind quality enhancement (BQE) model for compressed dynamic point clouds. BQE enhances compressed point clouds under unknown distortion levels by exploiting temporal dependencies and jointly modeling feature similarity and differences across multiple distortion levels. It consists of a joint progressive feature extraction branch and an adaptive feature fusion branch. In the joint progressive feature extraction branch, consecutive reconstructed frames are first fed into a recoloring-based motion compensation module to generate temporally aligned virtual reference frames. These frames are then fused by a temporal correlation-guided cross-attention module and processed by a progressive feature extraction module to obtain hierarchical features at different distortion levels. In the adaptive feature fusion branch, the current reconstructed frame is input to a quality estimation module to predict a weighting distribution that guides the adaptive weighted fusion of these hierarchical features. When applied to the latest geometry-based point cloud compression (G-PCC) reference software, i.e., test model category13 version 28, BQE achieved average PSNR improvements of 0.535 dB, 0.403 dB, and 0.453 dB, with BD-rates of -17.4\%, -20.5\%, and -20.1\% for the Luma, Cb, and Cr components, respectively.
\end{abstract}


\section{Introduction}
\IEEEPARstart{P}{oint} clouds represent objects or scenes as sets of 3D points, each with geometric coordinates and associated attributes \cite{z1,z2,z3,z4}. Owing to their ability to capture fine-grained spatial details, point clouds have been widely applied in fields such as autonomous driving, virtual reality, online education, and cultural heritage preservation \cite{z5,z6,z7,z8,z9,z10}. However, their massive data volume poses a significant challenge for efficient storage and transmission. To address this challenge, point cloud compression has been extensively studied \cite{z11,z12}. In 2017, the Moving Picture Experts Group (MPEG) under ISO/IEC issued a call for proposals on point clouds compression \cite{z13}, which led to the development of the geometry-based point cloud compression (G-PCC) standard \cite{z14}. Despite these advances, lossy compression inevitably introduces irreversible compression artifacts, which significantly degrade the quality of reconstructed point clouds.

In recent years, many quality enhancement methods for compressed point clouds have been proposed. These methods are generally categorized into traditional approaches and deep learning–based approaches. Traditional methods are not data-driven but rely on linear model–based filtering techniques, such as Kalman and Wiener filters, whose parameters are derived from assumed spatial or temporal correlations and noise statistics. Although these approaches can achieve certain improvements, their reliance on prior models and assumptions limits their ability to handle diverse and complex noise patterns. For deep learning–based approaches, graph convolution \cite{z15,z16,z17,z18} and sparse convolution \cite{z19,z20,z21,z22} are usually adopted for efficient feature extraction. By modelling local geometric relationships on the adjacency graph or sparse voxels of the input point cloud, these methods achieve higher reconstruction quality by recovering fine structural and texture details. 

Although existing deep learning-based approaches have demonstrated excellent performance, they are non-blind quality enhancement approaches. They typically require training multiple models with the same architecture for different distortion levels, which are usually controlled by the quantization parameters (QPs) of the codec. These non-blind quality enhancement methods suffer from two major limitations. First, in transcoding and transmission, the exact distortion level is often incomplete or unknown, making it difficult to select an appropriate trained model. Second, handling different distortion levels with separate, structurally identical models increases deployment complexity and resource requirements, which poses challenges for real-world applications.

To address the above challenges, we propose the first blind quality enhancement (BQE) model for compressed point cloud attributes. The proposed BQE model enhances compressed point clouds under unknown distortion levels by exploiting temporal dependencies and jointly modeling feature similarity and discrepancy across different distortion levels. It consists of two branches: a joint progressive feature extraction branch and an adaptive feature fusion branch. Specifically, consecutive reconstructed frames are first fed into a recoloring-based motion compensation (RMC) module to generate temporally aligned virtual reference frames. Subsequently, a temporal correlation-guided cross-attention (TCCA) module is proposed to fuse multi-frame information and efficiently exploit temporal correlations. The fused features are then processed by a progressive feature extraction module to obtain hierarchical representations corresponding to different distortion levels. The adaptive feature fusion branch takes the current reconstructed frame as input and predicts a quality vector (i.e., a weighting distribution) across different distortion levels via a quality estimation (QE) module, which guides the fusion of the progressively extracted hierarchical features. In summary, the contributions of this paper are as follows.
\begin{itemize}
\item{We propose a blind quality enhancement model for point cloud attributes that achieves effective restoration under unknown distortion levels. To the best of our knowledge, this is the first attempt at blind quality enhancement for compressed point clouds.}
\item{We propose a joint progressive feature extraction branch and an adaptive feature fusion branch in the BQE model. The joint progressive feature extraction branch captures representations across different distortion levels, while the adaptive feature fusion branch predicts a weighting distribution to guide adaptive feature fusion. This design enables effective utilization of multi-level distortion characteristics and improves reconstruction quality.}
\item{We propose a temporal correlation-guided cross-attention module that constructs a temporal interaction mechanism centered on the current frame to adaptively capture inter-frame correlations and dynamically assign attention weights, efficiently exploiting temporal information for enhancement.}
\item{We design a neighborhood-aware attention (NA) module that uses local positional encoding to strengthen geometry priors, enabling more efficient feature extraction for progressive feature extraction and weighting distribution prediction.}
\end{itemize}

The remainder of this paper is organized as follows. Section II reviews related work. Section III describes the proposed method. Section IV presents experimental results and analysis. Finally, Section V concludes the paper.

\section{Related Work}
In this section, we first review relevant work on non-blind point cloud quality enhancement. Then, because blind quality enhancement for point clouds has not been previously explored, we summarize representative blind quality enhancement methods for images and videos.
\subsection{Non-blind point cloud quality enhancement}
Non-blind point cloud quality enhancement methods can be roughly divided into two classes: traditional methods and deep learning-based methods.

Early studies introduced the Kalman filter into G-PCC to enhance attribute reconstruction. However, because its effectiveness relies on the stationarity assumption, the improvements were mainly observed in chrominance components \cite{a1}. To further mitigate distortion accumulation during the coding process, Wiener filter-based methods were subsequently proposed \cite{a2,a3}. Although these approaches offer some advantages, their performance remains constrained by the undelying assumption of linear distortion.

Deep learning-based methods can be broadly categorized into two categories: graph convolution-based methods and sparse convolution-based methods. Graph convolution-based methods use geometry priors to guide attribute enhancement. Sheng et al.  \cite{a4} proposed a multi-scale graph attention network to remove attribute artifacts caused by the G-PCC encoder. They constructed a graph based on geometry coordinates and used Chebyshev graph convolution to extract attribute feature representations. Xing et al. \cite{a5} introduced a graph-based quality enhancement network that uses geometry information as an auxiliary input and graph convolution blocks to extract local features efficiently. In \cite{a6}, we proposed PCE-GAN, which models point cloud attribute quality enhancement as an optimal transport problem and explicitly integrates perceptual quality into the enhancement framework. By using sparse 3D convolution, Liu et al. \cite{a7} proposed a dynamic point cloud enhancement method that uses inter-frame motion prediction with relative positional encoding and motion consistency to align the current frame with its reference. Zhang et al. \cite{a8} proposed G-PCC++, which enhances the quality of both geometry and attributes. The method densifies the decoded geometry via linear interpolation to form a continuous surface, applies Gaussian distance-weighted mapping for recoloring, and further refines the results with an attribute enhancement model. Later, they \cite{a9} proposed a fully data-driven method and a rule-unrolling-based optimization to restore G-PCC compressed point cloud attributes. Moreover, they \cite{a10} designed a learning-based adaptive in-loop filter for efficient quality enhancement. 

However, all the existing methods require prior knowledge of distortion levels and separate training of multiple architecture-identical models for different distortion levels, which results in high computational cost and limited applicability in real-world scenarios. Therefore, developing blind point cloud quality enhancement methods has become an urgent necessity.
\subsection{Blind image/video quality enhancement}
Although the data structure of images, videos, and point clouds are different, blind quality enhancement methods for compressed images and videos can still inspire the design of blind quality enhancement for compressed point cloud attributes. 

To remove compression artifacts, Kim et al. \cite{b1} proposed a pseudo-blind convolutional neural network (PBCNN) that estimates the compression quality factor and uses several associated non-blind models to remove compression artifacts for both video and image coding standards \cite{JPEG,H264}. To reduce computational costs, Xing et al. \cite{b2} proposed a resource-efficient blind quality enhancement (RBQE) method for compressed images. Leveraging dynamic deep neural networks and no-reference quality assessment methods, RBQE enables blind and progressive enhancement with an early-exit strategy, making it highly resource-friendly. Jiang et al. \cite{b3} proposed a flexible blind CNN for JPEG artifact removal that decouples and predicts an adjustable quality factor from the input image and injects it via a quality-factor attention block to control the trade-off between artifact removal and detail preservation. Xing et al. \cite{b5} proposed a blind image compression artifact reduction recurrent network that adapts to unknown quality factors and different degradation levels with scalable recurrent convolution and efficient convolution groups. Recently, Li et al. \cite{b6,b7} proposed a prompt-learning-based compressed image restoration network, which uses prompt learning to implicitly encode compression information and provide dynamic content-aware and distortion-aware guidance for restoration. Finally, Ding et al. \cite{b4} proposed a blind quality enhancement method for compressed video that exploits fluctuating temporal information and feature correlations across multiple QPs.
\begin{figure*}
\centering
\includegraphics[width=6.8in]{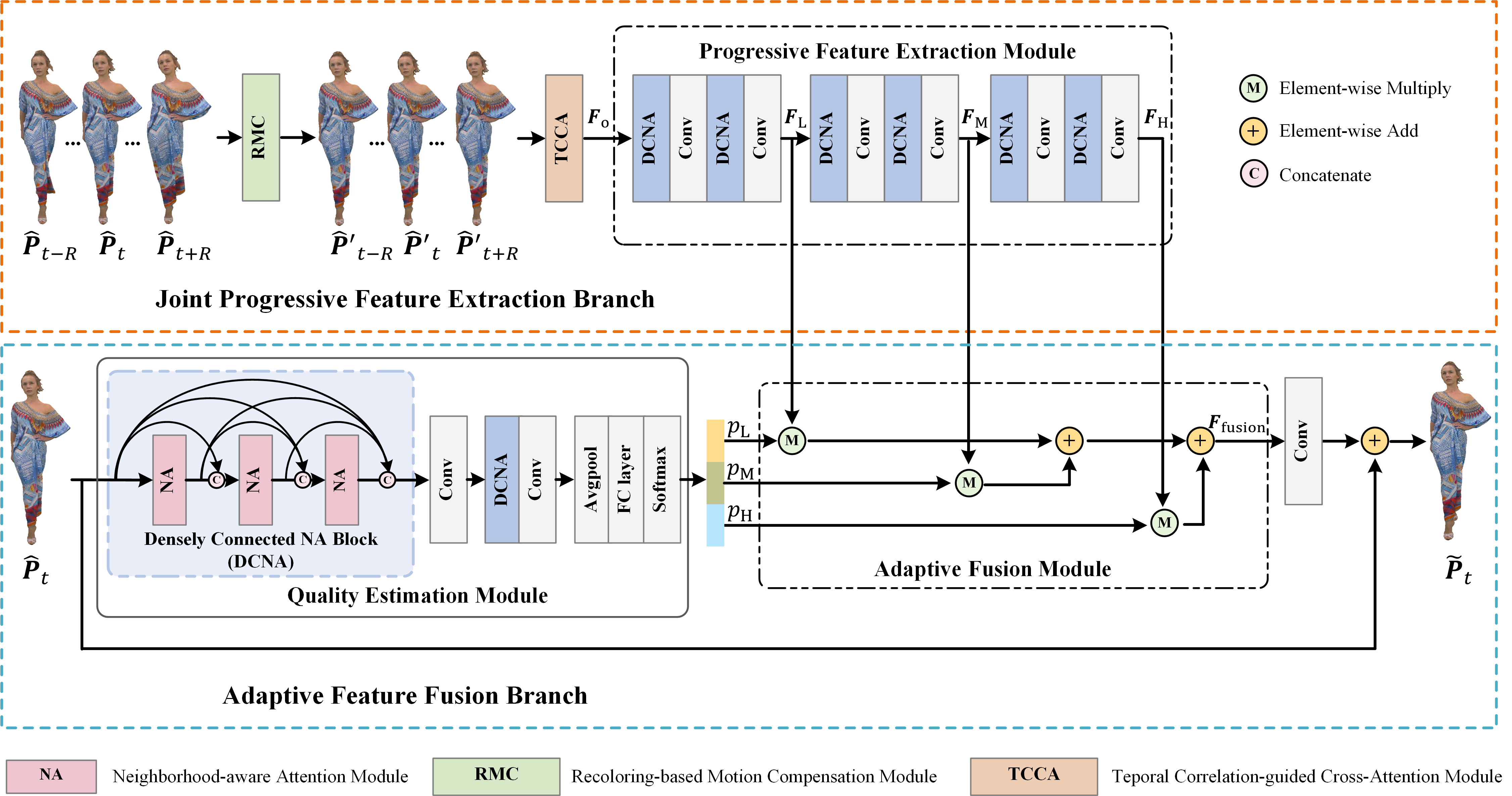}
\caption{BQE architecture. The proposed BQE model consists of two branches: a joint progressive feature extraction branch and an adaptive feature fusion branch. Given a reconstructed point cloud sequence $(\hat{\bm{P}}_{t-R}, \ldots, \hat{\bm{P}}_{t}, \ldots, \hat{\bm{P}}_{t+R})$, where \(\hat{\bm{P}}_{t}\) is the target frame and the remaining frames are reference frames, the goal of blind attribute quality enhancement is to restore \(\hat{\bm{P}}_{t}\) to an enhanced version \({\widetilde{\bm{P}}}_t\) without knowing the distortion level of \(\hat{\bm{P}}_{t}\). Note that the target frame is unchanged by RMC, i.e., $\hat{\bm{P}}'_t = \hat{\bm{P}}_t$.}
\label{FIG1}
\end{figure*}

\section{Proposed Method}
Given a point cloud sequence $\bm{P}$, the $t$-th frame is denoted by $\bm{P}_t = [\bm{P}_t^{G}, \bm{P}_t^{A}]$, where $\bm{P}_t^{G}$ represents the geometry and $\bm{P}_t^{A}$ represents the associated attributes. Let $\hat{\bm{P}}_t = [\hat{\bm{P}}_t^{G}, \hat{\bm{P}}_t^{A}]$ be the reconstructed frame after compression. For a temporal window of radius $R\in\mathbb{N}$, we consider the ordered window $(\hat{\bm{P}}_{t-R},\,\ldots,\,\hat{\bm{P}}_t,\,\ldots,\,\hat{\bm{P}}_{t+R})$, which has length $2R+1$. We assume lossless geometry compression, i.e., $\hat{\bm{P}}_t^{G} = \bm{P}_t^{G}$, so any distortion introduced by compression is restricted to the attributes $\hat{\bm{P}}_t^{A}$. The goal of attribute quality enhancement is to refine the reconstructed point cloud sequence within the temporal window to obtain higher-quality attributes, using the corresponding original sequence as ground truth during training (the original point clouds are not available at inference). In particular, frame $\hat{\bm{P}}_t$ is treated as the \emph{target} frame, while the remaining frames
in the window serve as \emph{reference} frames. 

The enhanced target frame is defined as $\tilde{\bm{P}}_t = [\bm{P}_t^{G},\,\tilde{\bm{P}}_t^{A}]$, which preserves the geometry and refines only the attributes. We compute
\begin{equation}
\tilde{\bm{P}}_t = \Psi_{\text{BQE}}( \hat{\bm{P}}_{t-R}, \ldots, \hat{\bm{P}}_t, \ldots, \hat{\bm{P}}_{t+R}; \boldsymbol{\theta} ),
\end{equation}
where \(\Psi_{\text{BQE}}(\cdot)\) denotes the proposed BQE model with learnable parameters \(\boldsymbol{\theta}\). In our experiments, we set \(R=2\). 

The proposed BQE model consists of a joint progressive feature extraction branch and an adaptive feature fusion branch (Fig. 1). Given the ordered input window $(\hat{\bm{P}}_{t-R},\,\ldots,\,\hat{\bm{P}}_t,\,\ldots,\,\hat{\bm{P}}_{t+R})$, the joint branch processes the entire window, whereas the fusion branch processes the center frame $\hat{\bm{P}}_t$. In the joint branch (Section III-A), the frames $(\hat{\bm{P}}_{t-R},\,\ldots,\,\hat{\bm{P}}_{t+R})$ are first passed through the RMC module to generate virtual reference frames $(\hat{\bm{P}}'_{t-R},\,\ldots,\,\hat{\bm{P}}'_t,\,\ldots,\,\hat{\bm{P}}'_{t+R})$, which align the reference frames to the target frame $\hat{\bm{P}}_t$. The aligned frames are then fed into the TCCA module, which models temporal correlations via multi-frame feature interaction and outputs a fused feature \(\bm{F}_o\). Next, \(\bm{F}_o\) is forwarded to the progressive feature extraction module, which extracts hierarchical features, enabling resource-efficient utilization of feature similarity across multiple distortion levels. In the adaptive feature fusion branch (Section III-B), the current reconstructed frame \(\hat{\bm{P}}_{t}\) is first processed by the QE module to estimate a quality vector that represents weighting distributions for features. The estimated quality vector is then fed into the adaptive fusion module to combine the progressively extracted hierarchical features.
\begin{figure}
\centering
\includegraphics[width=3.3in]{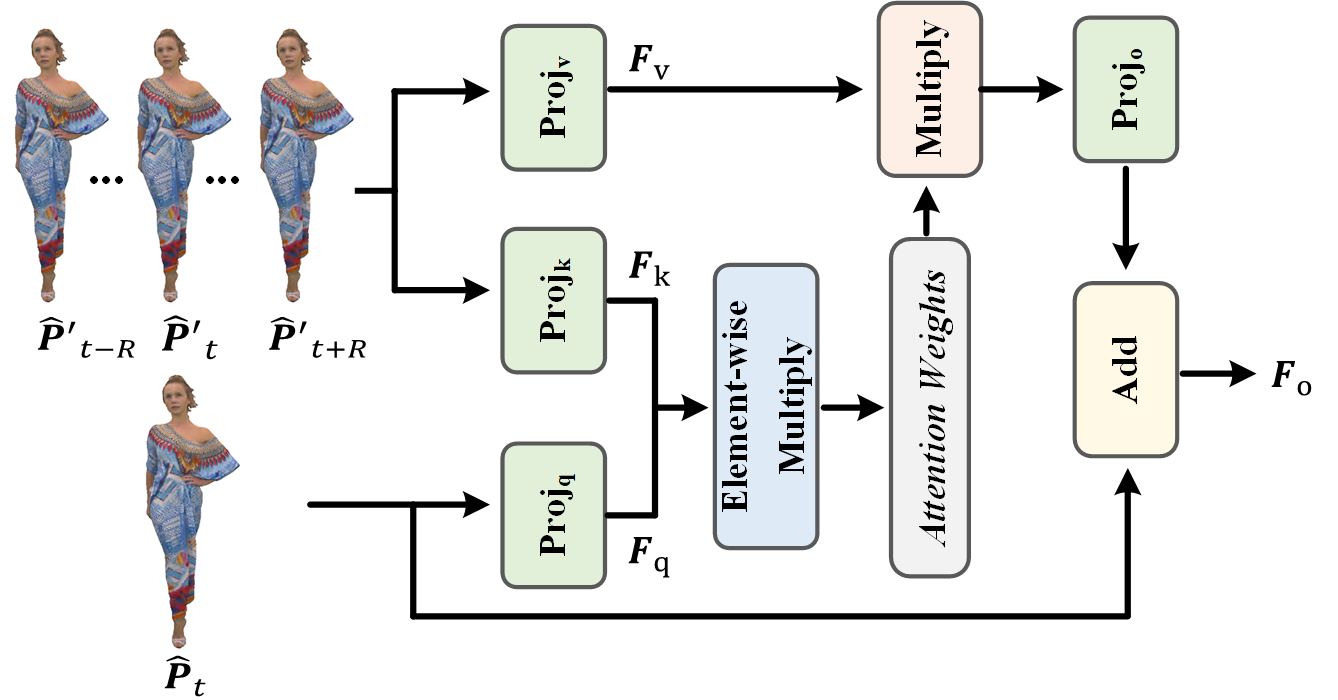}
\caption{Structure of the TCCA module. The ordered sequence $(\hat{\bm{P}}'_{t-R},\,\ldots,\,\hat{\bm{P}}'_{t},\,\ldots,\,\hat{\bm{P}}'_{t+R})$ denotes the virtual frames produced by the RMC module. For notational consistency, we use a prime to indicate frames after RMC. In particular, the target frame is unchanged by RMC, i.e., $\hat{\bm{P}}'_{t}=\hat{\bm{P}}_{t}$.}
\label{FIG2}
\end{figure}

\subsection{Joint Progressive Feature Extraction Branch}
To efficiently exploit temporal correlations in compressed dynamic point clouds, consecutive reconstructed frames are processed jointly. The RMC module first maps the attribute information of reference frames onto the geometry of the current frame to achieve precise spatial alignment. Next, the TCCA module adaptively aggregates temporal information from the reference frames by enabling information exchange between the current-frame and reference-frame features, improving the fused representation.Next, the TCCA module adaptively aggregates temporal information from the reference frames by enabling information exchange between the current-frame and reference-frame features, improving the fused representation. Finally, the fused feature \(\bm{F}_o\) is refined by the progressive feature extraction module which learns hierarchical features across different distortion levels to achieve effective quality enhancement.

\noindent\textit{\textbf{1) Recoloring-based Motion Compensation Module}}

To address the geometry and color misalignment caused by inter-frame motion in dynamic point clouds, we use the RMC module. Instead of explicitly estimating motion vectors, RMC remaps the attribute information of the reference frames to the geometry coordinates of the current frame, achieving spatial alignment and reducing motion-induced artifacts. These virtual reference frames facilitate more accurate temporal feature fusion in subsequent modules.

\noindent\textit{\textbf{2) Temporal Correlation-guided Cross-Attention Module }}

To model temporal correlations, we propose a TCCA module that adaptively fuses information across multi-frame point clouds (Fig. 2). Specifically, the attribute of the current frame \({\hat{\bm{P}}}_t\) is first processed by a projection layer \(\mathrm{Proj}_\mathrm{q}\) to construct the query embedding. Meanwhile, the consecutive frames \(\{\hat{\bm{P}}'_{t-R}, \ldots, \hat{\bm{P}}'_t, \ldots, \hat{\bm{P}}'_{t+R}\}\) are fed into separate projection layer \(\mathrm{Proj}_\mathrm{k}\) and \(\mathrm{Proj}_\mathrm{v}\) to generate the key and value embeddings:
\begin{equation}
\begin{aligned}
\bm{F}_q &= \mathrm{Proj}_\mathrm{q}(\hat{\bm{P}}'^A_t), \\
\bm{F}_k &= \mathrm{Proj}_\mathrm{k}(\mathrm{Concat}(\hat{\bm{P}}'^A_i)), \\
\bm{F}_v &= \mathrm{Proj}_\mathrm{v}(\mathrm{Concat}(\hat{\bm{P}}'^A_i)).
\end{aligned}
\end{equation}
where \(i \in \{t - R, \ldots, t + R\}\). 

\textit{Note that} $\hat{\bm P}'_t = \hat{\bm P}_t$, since the target frame at time $t$ remains unchanged after the RMC module, and the prime symbol is introduced only for notational consistency.

\(\mathrm{Proj}_\mathrm{q}\), \(\mathrm{Proj}_\mathrm{k}\), and \(\mathrm{Proj}_\mathrm{v}\) refer to the projection layers that comprise cascaded linear layers and LeakyReLU functions, which can be mathematically represented as \(\mathrm{Linear}\big(
\mathrm{LReLU}\big(
\mathrm{Linear}\big(
\mathrm{LReLU}\big(
\mathrm{Linear}(\cdot)
\big)
\big)
\big)
\big)\).

Then, the similarity between \(\bm{F}_q\) and \(\bm{F}_k\) is computed using scaled dot-product attention, yielding the attention weights \(\bm{s}\) \cite{z23}:
\begin{equation}
\bm{s} = \mathrm{Softmax}(
\frac{\bm{F}_q \cdot \bm{F}_k^{\top}}{\sqrt{d_k}}),
\end{equation}
where $d_k$ denotes the embedding dimension of the keys (and queries). Next, the attention output is obtained by weighting the value embeddings, $s\bm{F}_v$, and projecting the result through a linear layer \(\mathrm{Proj}_\mathrm{o}\) to produce $\bm{F}_t$. Moreover, we introduce a skip connection from $\hat{\bm{P}}^A_t$ to preserve current-frame information. Finally, the resulting features are concatenated with the geometry of the current frame:
\begin{equation}
\bm{F}_o = \mathrm{Concat}(
\bm{F}_t + \hat{\bm{P}}^A_{t}, \bm{P}_t^{G}).
\end{equation}
The fused feature $\bm{F}_o$ is then forwarded to the progressive feature extraction module.

\noindent\textit{\textbf{3) Progressive Feature Extraction Module}}

The design of the progressive extraction module is primarily motivated by two key considerations.

\textcircled{1} \textbf{Feature similarity across multiple distortion levels:} Compressed point clouds encoded by different QPs share similar feature representations in the quality enhancement task. Therefore, the feature extraction structures corresponding to various distortion levels can be partially shared within a unified framework to improve efficiency and consistency.

\textcircled{2} \textbf{Context modeling requirements for high-distortion point clouds:} Point clouds with severe compression distortion often suffers from significant textural degradation. Hence, quality enhancement for such data requires a deeper network architecture to expand the receptive field and capture richer contextual dependencies. 

Based on the above considerations, the progressive feature extraction module is designed as a multi-stage structure with multiple cascaded densely connected NA (DCNA) blocks for hierarchical feature extraction. Specifically, a shallow network is used for lightly distorted point clouds to capture local features, whereas a deeper network is adopted for more severely distorted point clouds to extract richer contextual information. Moreover, the deeper configurations reuse the architecture of the shallower ones, enabling resource-efficient enhancement across distortion levels. As illustrated in Fig. 1, shallow, medium, and deep network structures are used for low distorted, medium distorted, and highly distorted point clouds, producing features denoted by \(\bm{F}_L\), \(\bm{F}_M\), \(\bm{F}_H\), respectively. Details of the NA module are provided in Section III-B.

\subsection{Adaptive Feature Fusion Branch}
To fully exploit the feature similarity and difference across multiple distortion levels and enable a unified quality enhancement functionality for compressed point clouds, we design a QE module to estimate the distortion level of the current frame and generate a corresponding quality vector. Guided by this quality vector, the progressively extracted hierarchical features \([\bm{F}_L, \bm{F}_M, \bm{F}_H ]\) are fused in the adaptive fusion module, achieving distortion-aware feature integration, as illustrated in Fig. 1.
\begin{figure}
\centering
\includegraphics[width=3.3in]{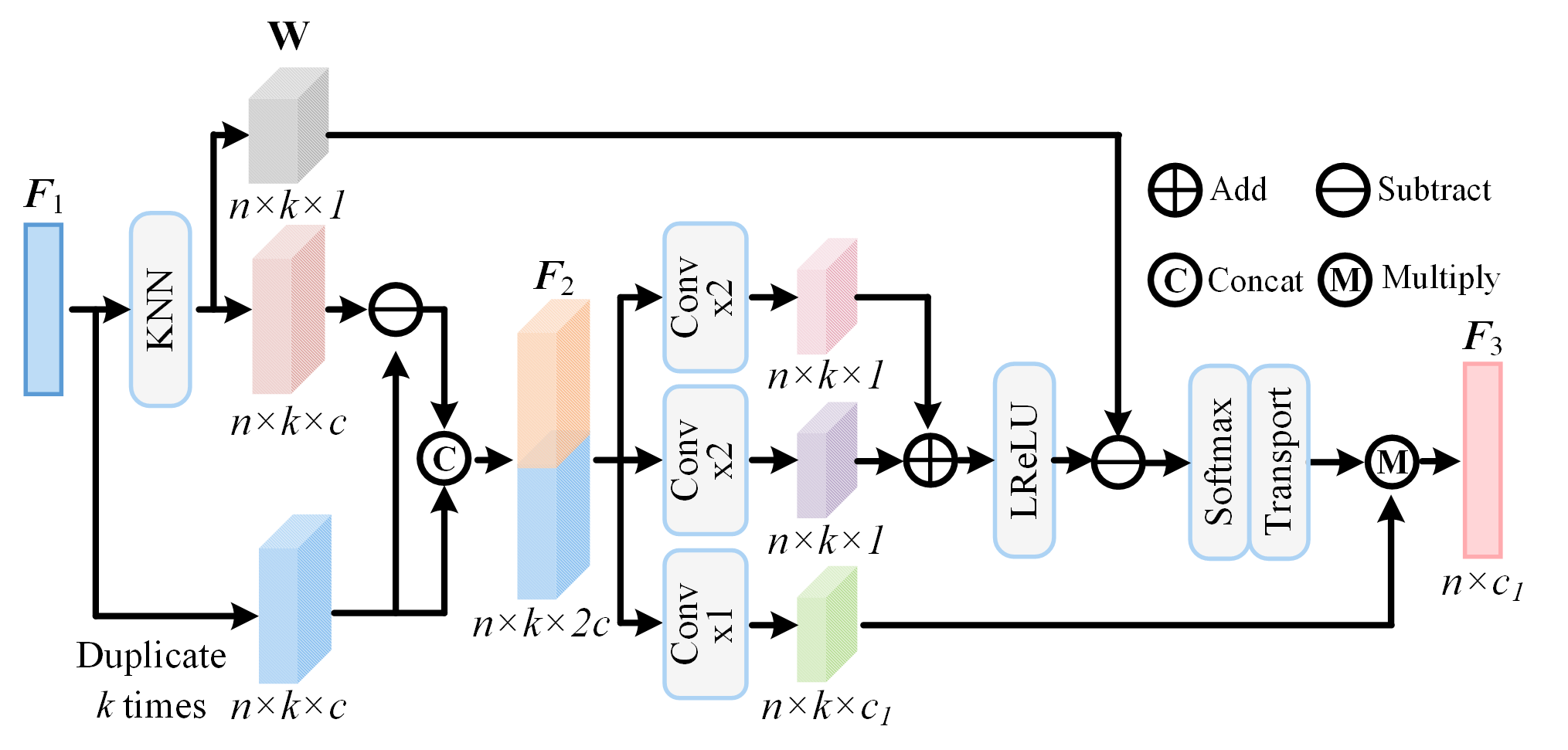}
\caption{Architecture of the NA module.}
\label{FIG3}
\end{figure}

\noindent\textit{\textbf{1) Neighborhood-Aware Attention Module}}

The NA module is illustrated in Fig. 3. It extracts geometry and attribute features jointly through dynamic neighborhood-based feature aggregation and adaptive weighting mechanisms. Specifically, the input feature \(\bm{F}_1\in\mathbb{R}^{n\times c}\), where \(n\) denotes the number of points, is first used to dynamically construct local neighborhoods via k-nearest neighbor (KNN) \cite{z24} search, establishing associations between each center point and its \(k\) neighboring points. The center point and its neighboring features are then concatenated to form a composite representation \(\bm{F}_2\in\mathbb{R}^{n\times k\times2c}\) that captures both geometry and texture dependencies. The composite feature \(\bm{F}_2\) is processed by two \(1\times1\) convolution layers followed by LeakyReLU activations to extract high-dimensional representations, while a distance matrix \(\bm{W}\in\mathbb{R}^{n\times k\times1} \)is incorporated for positional encoding to explicitly encode spatial relationships and enhance the representation of local geometry structures. Finally, a Softmax operation is used to assign attention weights based on feature similarity, enabling adaptive weighting and aggregation of important points to produce the output feature \(\bm{F}_3\in\mathbb{R}^{n\times c_1}\).

\noindent\textit{\textbf{2)	Quality Estimation Module}}

As illustrated in Fig. 1, the QE module consists of densely connected NA modules and \(1\times1\) convolution layers. In addition, an average pooling layer, a fully connected layer, and a softmax activation function are used to generate the final estimated quality vector \(\bm{p} = [p_L, p_M, p_H]^{\top}\), representing the probabilities of the current frame belonging to low- distortion, medium-distortion, and high-distortion levels, respectively, where \(p_L+p_M+p_H=1\). This vector serves as a guiding weight in the subsequent adaptive fusion module for feature integration across different distortion levels.

\noindent\textit{\textbf{3)	Adaptive Fusion Module}}

Since features across different distortion levels exhibit both similarity and difference, distortion-level adaptive feature fusion plays a crucial role in blind quality enhancement. On one hand, point clouds at low-distortion, medium-distortion, and high-distortion levels share certain commonalities in spatial structure and basic attribute distribution, where features from lower distortion levels can provide valuable priors for enhancing highly distorted samples. On the other hand, notable differences exist among these levels, thus applying identical enhancement strategies to all of them may lead to suboptimal performance.

Therefore, as illustrated in Fig. 1, BQE uses a quality-vector-based adaptive weighting mechanism to dynamically fuse features from three distortion levels. Specifically, each probability \([p_L, p_M, p_H]^{\top}\) is multiplied by its corresponding feature \([\bm{F}_L, \bm{F}_M, \bm{F}_H ]\), and the three weighted features are summed to obtain the fused feature:
\begin{equation}
\label{8}
\bm{F}_{\mathrm{fusion}} = \sum_{i \in \{L, M, H\}} p_i \bm{F}_i.
\end{equation}
\subsection{Loss Function}
To reduce the training burden and accelerate convergence, we first pre-train the QE module and then train the overall BQE end to end while keeping the QE parameters fixed.

The QE module is pre-trained using a cross-entropy loss. However, since the low-distortion, medium-distortion, and high-distortion levels have a clear ordinal relationship, one-hot labels ignore this structure. To address this, we construct Gaussian kernel-based, quality-aware soft labels for each reconstructed frame. Specifically, point clouds compressed at six bitrates (denoted R01, R02, R03, R04, R05, and R06 with corresponding QPs \({QP}_{R01}=51,\ {QP}_{R02}=46,\ {QP}_{R03}=40,\ {QP}_{R04}=34,\ {QP}_{R05}=28,\ {QP}_{R06}=22\)) are grouped into three distortion levels according to degradation severity: R01 and R02 correspond to high-distortion, R03 and R04 correspond to medium-distortion, and R05 and R06 correspond to low-distortion. The QP centers for the three distortion levels are defined as
\begin{equation}
\left\{
\begin{aligned}
c_H &= \frac{QP_{R01} + QP_{R02}}{2}, \\
c_M &= \frac{QP_{R03} + QP_{R04}}{2}, \\
c_L &= \frac{QP_{R05} + QP_{R06}}{2}.
\end{aligned}
\right.
\end{equation}

For a frame with QP of \(q\), we construct a quality-aware soft label vector \(\bm{g}=[g_L, g_M, g_H]\) using a Gaussian kernel:
\begin{equation}
g_i =
\frac{
\exp(
-\frac{1}{2}(\frac{q - c_i}{\sigma})^2
)
}{
\sum_{j \in \{L, M, H\}}
\exp(
-\frac{1}{2}(\frac{q - c_j}{\sigma})^2
)
},
\end{equation}
where \(i\in\{L,M,H\}\), and \(\sigma\) denotes a hyperparameter controlling the smoothness of the labels, which is set to 5 in our experiments. The resulting \(\bm{g}\) satisfies \(g_L+g_M+g_H=1\) and numerically characterizes the relative proximity of the current frame to the three distortion levels. Accordingly, the loss function of the QE module can be expressed as
\begin{equation}
\begin{aligned}
\mathcal{L}_{\mathrm{QE}}(\boldsymbol{\theta}_{QE})
= - \sum_{i \in \{L, M, H\}} g_i \log(p_i),
\end{aligned}
\end{equation}
where \(\bm{\theta}_{QE}\) denotes the learnable parameters of the QE module, and \(\Psi_{QE}(\cdot)\) represents the mapping function of the module, and \(\bm{p}=\Psi_{\mathrm{QE}}(\hat{\bm{P}}_t; \boldsymbol{\theta}_{{QE}})\) with \(\sum_{i \in \{L, M, H\}} p_i=1\). By minimizing \(\mathcal{L}_{\mathrm{QE}}\), the QE module learns to generate a quality-aware distortion-level probability vector, which serves as an adaptive weighting prior to guide feature fusion across different distortion levels in the subsequent module.

Then, the overall BQE model is trained with the fixed parameters \(\bm{\theta}_{QE}\). The objective of this stage is to minimize the mean squared error (MSE) between the enhanced attributes \({\tilde{\bm{P}}}^A_t\) and the original attributes \(\bm{P}^A_t\):
\begin{equation}
\begin{aligned}
\mathcal{L}_{\mathrm{BQE}}(\boldsymbol{\theta}_{{BQE-QE}})
= \frac{1}{n} \left\| \tilde{\bm{P}}^A_t - \bm{P}^A_t \right\|_2^2.
\end{aligned}
\end{equation}
where $\tilde{\bm{P}}^A_t=\Psi_{\mathrm{BQE}}(\hat{\bm{P}}^A_{t-R},\ldots,\hat{\bm{P}}^A_{t+R};\boldsymbol{\theta}_{\mathrm{BQE\text{-}QE}})$, \(n\) denotes the number of points, and \(\bm{\theta}_{BQE-QE}\) indicates the parameters of the BQE model excluding those of the QE module. 

\section{Experimental Results and Analysis}
This section presents our experimental results. Specifically, Section IV-A describes the experimental setup, including the datasets, implementation details, and evaluation metrics. Section IV-B provides objective evaluations of the enhanced point clouds and compares coding efficiency before and after integrating the proposed method into the G-PCC-based compression framework. Section IV-C examines the robustness of the proposed BQE model. Section IV-D compares BQE with several state-of-the-art deep learning–based point cloud quality enhancement methods. Section IV-E shows visual comparisons between BQE and the other methods. Section IV-F analyzes the computational complexity of BQE. Finally Section IV-G presents ablation studies to investigate the contribution of each component to the overall performance.
\subsection{Experimental Setup}
\noindent\textit{\textbf{1) Datasets}}

We trained the proposed model using five dynamic point cloud sequences: \textit{Longdress}, \textit{Basketball}, \textit{Exercise}, \textit{Andrew}, and \textit{David}. \textit{Longdress} comes from the 8i Voxelized Full Bodies dataset (8iVFB v2) \cite{b20} with 10-bit precision. \textit{Basketball} and \textit{Exercise} come from the Owlii Dynamic Human Textured Mesh Sequence dataset (Owlii) \cite{b21} with 11-bit precision. \textit{Andrew} and \textit{David} come from the Microsoft Voxelized Upper Bodies dataset (MVUB) \cite{b22} with 10-bit precision. The frame rate of each sequence is 30 fps. We encoded the sequences using the G-PCC test model, TMC13v28 \cite{b25}, applying inter-frame prediction with an octree-RAHT configuration to generate training datasets. We conducted the encoding under the Common Test Condition (CTC) of C1 \cite{b24}, which involves lossless geometry compression and lossy attribute compression. We collected the first 32 frames of each sequence for training, for a total of 160 frames. Due to limitations in GPU memory capacity, we used a patch generation-and-fusion approach as in \cite{a5}.
\begin{table}[ht]
\centering
\caption{\(\Delta\)PSNR (dB) and BD-rate (\%) after integrating BQE into G-PCC}
\label{tab:my_label1}  
\resizebox{8.55cm}{!}{
\begin{tabular}{lcccccccc}
\toprule
\multirow{2}{*}{\textbf{Sequence}} & \multicolumn{4}{c}{\textbf{\(\Delta\)PSNR (dB)}} & \multicolumn{4}{c}{\textbf{BD-rate (\%)}} \\ 
 & \textbf{Y} & \textbf{Cb} & \textbf{Cr} & \textbf{YCbCr} & \textbf{Y} & \textbf{Cb} & \textbf{Cr} & \textbf{YCbCr} \\ \hline
        Dancer & 0.461 & 0.153 & 0.254 & 0.396 & -14.1 & -9.8 & -12.4 & -13.3 \\ 
        Model & 0.443 & 0.221 & 0.422 & 0.413 & -14.9 & -15.0 & -19.7 & -15.5 \\ 
        Redandblack & 0.512 & 0.435 & 0.508 & 0.502 & -16.7 & -21.3 & -14.7 & -17.0 \\ 
        Soldier & 0.581 & 0.485 & 0.474 & 0.556 & -19.9 & -28.6 & -25.9 & -21.7 \\ 
        Loot & 0.428 & 0.764 & 0.590 & 0.490 & -17.1 & -35.6 & -28.3 & -20.8 \\ 
        Queen & 0.513 & 0.532 & 0.615 & 0.528 & -16.8 & -19.8 & -20.7 & -17.6 \\ 
        Phil & 0.541 & 0.229 & 0.256 & 0.466 & -16.3 & -12.5 & -12.4 & -15.3 \\ 
        Ricardo & 0.634 & 0.376 & 0.520 & 0.588 & -20.6 & -22.3 & -26.5 & -21.5 \\ 
        Sarah & 0.705 & 0.433 & 0.437 & 0.638 & -20.7 & -19.3 & -20.0 & -20.4 \\ \hline
        Average & \textbf{0.535} & \textbf{0.403} & \textbf{0.453} & \textbf{0.508} & \textbf{-17.4} & \textbf{-20.5} & \textbf{-20.1} & \textbf{-18.1} \\ \bottomrule
\end{tabular}}
\end{table}
\begin{table}
\centering
\caption{\(\Delta\)PSNR (dB) of Y Component caused by BQE}
\label{tab:my_label3}  
\resizebox{8cm}{!}{
\begin{tabular}{lcccccc}
\toprule
\multirow{2}{*}{\textbf{Sequence}} & \multicolumn{6}{c}{\textbf{\(\Delta\)PSNR (dB)}} \\ 
 & \textbf{R01} & \textbf{R02} & \textbf{R03} & \textbf{R04} & \textbf{R05} & \textbf{R06}  \\ \hline
        Dancer & 0.523 & 0.596 & 0.603 & 0.544 & 0.353 & 0.145 \\ 
        Model & 0.387 & 0.505 & 0.600 & 0.617 & 0.483 & 0.067 \\ 
        Redandblack & 0.394 & 0.505 & 0.586 & 0.620 & 0.589 & 0.380 \\ 
        Soldier & 0.221 & 0.314 & 0.497 & 0.726 & 0.909 & 0.819 \\ 
        Loot & 0.300 & 0.335 & 0.384 & 0.553 & 0.643 & 0.351 \\ 
        Queen & 0.605 & 0.650 & 0.554 & 0.515 & 0.560 & 0.193 \\ 
        Phil & 0.363 & 0.590 & 0.739 & 0.738 & 0.598 & 0.218 \\ 
        Ricardo & 0.572 & 0.657 & 0.761 & 0.740 & 0.640 & 0.435 \\ 
        Sarah & 0.768 & 0.852 & 0.868 & 0.783 & 0.659 & 0.302 \\ \hline
        \textbf{Average} & \textbf{0.459} & \textbf{0.556} & \textbf{0.622} & \textbf{0.648} & \textbf{0.604} & \textbf{0.323} \\ \bottomrule
\end{tabular}}
\end{table}

We tested the performance of BQE on nine sequences: \textit{Loot}, \textit{Redandblack}, \textit{Soldier}, \textit{Dancer}, \textit{Model}, \textit{Phil}, \textit{Ricardo}, \textit{Sarah}, and \textit{Queen}. \textit{Loot}, \textit{Redandblack}, and \textit{Soldier} come from the 8iVFB v2 dataset with 10-bit precision. \textit{Dancer} and \textit{Model} come from the Owlii dataset with 11-bit precision. \textit{Phil}, \textit{Ricardo}, and \textit{Sarah} come from the MVUB dataset with 10-bit precision. \textit{Queen} comes from the Technicolor dataset \cite{b23} with 10-bit precision. The frame rate of Queen is 50 fps, while all other sequences have a frame rate of 30 fps. Each sequence was compressed using TMC13v28 with QPs 51, 46, 40, 34, 28, 22, corresponding to the six bitrates, R01, R02, R03, R04, R05, and R06. We collected the first 32 frames of each sequence for testing, for a total of 288 frames.

\noindent\textit{\textbf{2) Implementation Details}}

We trained the proposed BQE model for 50 epochs with a batch size of 10, using the Adam optimizer \cite{b26} with a learning rate of 0.0001. The number of nearest neighbors in the KNN algorithm was set to k\ =\ 20. The model was implemented in PyTorch v2.7 and trained on an NVIDIA GeForce RTX5090 GPU. We trained three models corresponding to the Y, Cb, and Cr color components, where each model was jointly trained across six bitrate levels (R01–R06) rather than trained separately for each bitrate.

\noindent\textit{\textbf{3) Evaluation Metrics}}

We evaluated the performance of the proposed BQE model using delta peak signal-to-noise ratio (\(\Delta\)PSNR) and BD-rate metrics \cite{b27}. The \(\Delta\)PSNR measures the PSNR difference between the proposed method and the anchor at a single bitrate while the BD-rate measures the average bitrate increment in bits per input point (bpip) at the same PSNR when integrating BQE into G-PCC. A positive \(\Delta\)PSNR and a negative BD-rate indicate performance gains achieved by the proposed method. In addition to calculating the PSNR for all the color components, we also used the YCbCr-PSNR, a weighted average of Y, Cb, and Cr PSNRs with a ratio of 6:1:1, to comprehensively evaluate the overall color quality gains achieved by the proposed method. 
\begin{table}
\centering
\caption{\(\Delta\)PSNR (dB) and BD-rate (\%) after integrating BQE into GeSTMv8}
\label{tab:my_label1}  
\resizebox{8.55cm}{!}{
\begin{tabular}{lcccccccc}
\toprule
\multirow{2}{*}{\textbf{Sequence}} & \multicolumn{4}{c}{\textbf{\(\Delta\)PSNR (dB)}} & \multicolumn{4}{c}{\textbf{BD-rate (\%)}} \\ 
 & \textbf{Y} & \textbf{Cb} & \textbf{Cr} & \textbf{YCbCr} & \textbf{Y} & \textbf{Cb} & \textbf{Cr} & \textbf{YCbCr} \\ \hline
        Dancer & 0.392 & 0.169 & 0.206 & 0.341 & -11.5 & -13.2 & -10.5 & -11.6 \\ 
        Model & 0.344 & 0.198 & 0.239 & 0.313 & -10.4 & -14.4 & -11.6 & -11.0 \\ 
        Redandblack & 0.401 & 0.332 & 0.328 & 0.384 & -15.2 & -18.3 & -9.5 & -14.9 \\ 
        Soldier & 0.422 & 0.361 & 0.294 & 0.398 & -13.0 & -21.0 & -16.3 & -14.4 \\ 
        Loot & 0.375 & 0.477 & 0.376 & 0.388 & -13.5 & -22.5 & -17.8 & -15.2 \\ 
        Queen & 0.442 & 0.367 & 0.415 & 0.429 & -14.1 & -15.2 & -16.1 & -14.5 \\ 
        Phil & 0.482 & 0.237 & 0.250 & 0.422 & -13.2 & -13.8 & -12.6 & -13.2 \\ 
        Ricardo & 0.491 & 0.189 & 0.315 & 0.431 & -13.1 & -10.8 & -15.3 & -13.1 \\ 
        Sarah & 0.530 & 0.175 & 0.266 & 0.452 & -14.0 & -8.2 & -11.7 & -13.0 \\ \hline
        Average & \textbf{0.431} & \textbf{0.278} & \textbf{0.299} & \textbf{0.395} & \textbf{-13.1} & \textbf{-15.3} & \textbf{-13.5} & \textbf{-13.4} \\ \bottomrule
\end{tabular}}
\end{table}

\begin{table}
\centering
\caption{\(\Delta\)PSNR (dB) of Y Component caused by BQE after integrating it into GeSTMv8}
\label{tab:my_label3}  
\resizebox{8cm}{!}{
\begin{tabular}{lcccccc}
\toprule
\multirow{2}{*}{\textbf{Sequence}} & \multicolumn{6}{c}{\textbf{\(\Delta\)PSNR (dB)}} \\ 
 & \textbf{R01} & \textbf{R02} & \textbf{R03} & \textbf{R04} & \textbf{R05} & \textbf{R06}  \\ \hline
        Dancer & 0.331 & 0.462 & 0.516 & 0.518 & 0.371 & 0.157 \\ 
        Model & 0.236 & 0.323 & 0.433 & 0.494 & 0.459 & 0.121 \\ 
        Redandblack & 0.233 & 0.359 & 0.466 & 0.509 & 0.517 & 0.324 \\ 
        Soldier & 0.203 & 0.247 & 0.422 & 0.491 & 0.605 & 0.562 \\ 
        Loot & 0.287 & 0.309 & 0.320 & 0.457 & 0.557 & 0.323 \\ 
        Queen & 0.525 & 0.596 & 0.526 & 0.470 & 0.440 & 0.095 \\ 
        Phil & 0.334 & 0.468 & 0.621 & 0.689 & 0.587 & 0.193 \\ 
        Ricardo & 0.432 & 0.508 & 0.598 & 0.626 & 0.536 & 0.245 \\ 
        Sarah & 0.457 & 0.622 & 0.619 & 0.664 & 0.555 & 0.261 \\ \hline
        \textbf{Average} & \textbf{0.338} & \textbf{0.433} & \textbf{0.502} & \textbf{0.546} & \textbf{0.514} & \textbf{0.253} \\ \bottomrule
\end{tabular}}
\end{table}
\subsection{Objective Quality Evaluation}

Table I reports the overall performance of BQE in terms of \(\Delta\)PSNR and BD-rate, averaged over the first 32 frames of each test sequence. The results show consistent quality gains across all color components, as well as clear improvements in the combined YCbCr evaluation. Table II further analyzes the performance of BQE on the Y component across different bitrates. The gains were observed at all bitrate settings and were particularly pronounced at medium bitrates (R03 and R04), indicating that BQE is especially effective in this operating range. Moreover, the rate-PSNR curves shown in Fig. 4 illustrate that integrating BQE into G-PCC consistently improved coding efficiency over a wide range of bitrates.

\subsection{Robustness Analysis}
To further assess the effectiveness and generalization of BQE, we conducted zero-shot evaluations (i.e., without retraining or fine-tuning) under different codecs and QPs settings. 

\noindent\textit{\textbf{1)	Application to Solid G-PCC codec}}

We integrated the trained BQE models into Solid G-PCC, a 3D point cloud compression standard under development, and evaluated them using its test platform, GeSTMv8 \cite{b28}. We compressed all test sequences using GeSTMv8 with inter-frame prediction and an octree-RAHT configuration. Table III reports the average \(\Delta\)PSNRs and BD-rates and shows consistent gains across the Y, Cb, and Cr components, as well as in the combined YCbCr evaluation. Table IV reports the \(\Delta\)PSNR results of the Y component across all six bitrates; the gains were greater at medium and high bitrates, consistent with Table II. Fig. 5 compares the rate-PSNR curves before and after integrating BQE into GeSTMv8. It shows that BQE also adapted to the Solid G-PCC encoder and improved coding efficiency.

\begin{table}
\centering
\caption{\(\Delta\)PSNR (dB) and BD-rate (\%) after integrating BQE into G-PCC under modified QPs}
\label{tab:my_label1}  
\resizebox{8.55cm}{!}{
\begin{tabular}{lcccccccc}
\toprule
\multirow{2}{*}{\textbf{Sequence}} & \multicolumn{4}{c}{\textbf{\(\Delta\)PSNR (dB)}} & \multicolumn{4}{c}{\textbf{BD-rate (\%)}} \\ 
 & \textbf{Y} & \textbf{Cb} & \textbf{Cr} & \textbf{YCbCr} & \textbf{Y} & \textbf{Cb} & \textbf{Cr} & \textbf{YCbCr} \\ \hline
        Dancer & 0.556 & 0.173 & 0.202 & 0.464 & -18.3 & -8.0 & -9.5 & -15.9 \\ 
        Model & 0.419 & 0.223 & 0.300 & 0.380 & -17.7 & -18.5 & -16.8 & -17.7 \\ 
        Redandblack & 0.479 & 0.433 & 0.534 & 0.480 & -19.6 & -20.9 & -10.5 & -18.6 \\ 
        Soldier & 0.455 & 0.426 & 0.357 & 0.439 & -14.4 & -26.2 & -18.1 & -16.3 \\ 
        Loot & 0.366 & 0.617 & 0.491 & 0.413 & -18.0 & -30.6 & -22.6 & -20.1 \\ 
        Queen & 0.582 & 0.477 & 0.622 & 0.574 & -13.1 & -10.3 & -13.4 & -12.8 \\ 
        Phil & 0.517 & 0.261 & 0.276 & 0.455 & -12.6 & -13.5 & -14.5 & -13.0 \\ 
        Ricardo & 0.601 & 0.311 & 0.484 & 0.550 & -20.6 & -20.6 & -22.9 & -20.9 \\ 
        Sarah & 0.682 & 0.351 & 0.317 & 0.595 & -11.9 & -7.2 & -5.4 & -10.5 \\ \hline
        Average & \textbf{0.517} & \textbf{0.364} & \textbf{0.398} & \textbf{0.483} & \textbf{-16.3} & \textbf{-17.3} & \textbf{-14.9} & \textbf{-16.2} \\ \bottomrule
\end{tabular}}
\end{table}

\begin{table}
\centering
\caption{\(\Delta\)PSNR (dB) OF Y Component caused by BQE after integrating it into G-PCC under modified QPs}
\label{tab:my_label3}  
\resizebox{8cm}{!}{
\begin{tabular}{lcccccc}
\toprule
\multirow{2}{*}{\textbf{Sequence}} & \multicolumn{6}{c}{\textbf{\(\Delta\)PSNR (dB)}} \\ 
 & \textbf{R01} & \textbf{R02} & \textbf{R03} & \textbf{R04} & \textbf{R05} & \textbf{R06}  \\ \hline
        Dancer & 0.547 & 0.590 & 0.638 & 0.618 & 0.562 & 0.383 \\ 
        Model & 0.357 & 0.407 & 0.486 & 0.499 & 0.457 & 0.308 \\ 
        Redandblack & 0.408 & 0.473 & 0.544 & 0.521 & 0.491 & 0.440 \\ 
        Soldier & 0.298 & 0.320 & 0.430 & 0.548 & 0.579 & 0.559 \\ 
        Loot & 0.303 & 0.311 & 0.315 & 0.350 & 0.445 & 0.471 \\ 
        Queen & 0.785 & 0.801 & 0.684 & 0.533 & 0.432 & 0.256 \\ 
        Phil & 0.351 & 0.446 & 0.649 & 0.698 & 0.614 & 0.342 \\ 
        Ricardo & 0.651 & 0.655 & 0.730 & 0.682 & 0.565 & 0.325 \\ 
        Sarah & 0.749 & 0.775 & 0.813 & 0.746 & 0.599 & 0.410 \\ \hline
        \textbf{Average} & \textbf{0.494} & \textbf{0.531} & \textbf{0.588} & \textbf{0.577} & \textbf{0.527} & \textbf{0.388} \\ \bottomrule
\end{tabular}}
\end{table}

\noindent\textit{\textbf{2)	Application to G-PCC under Modified QPs}}

We then replaced the original QP set \{51, 46, 40, 34, 28, 22\} with \{54, 49, 43, 37, 31, 25\} to evaluate the generalization ability of BQE. We compressed all test sequences using G-PCC under the new QP settings with inter-frame prediction and an octree-RAHT configuration. Table V reports the average PSNRs and BD-rates. BQE achieved average \(\Delta\)PSNR gains for the Y, Cb, and Cr components, and it also improved the combined YCbCr evaluation. Table VI reports the \(\Delta\)PSNR results of the Y component across all six bitrates. The gains were larger at medium and high bitrates, consistent with Table II. Fig. 6 compares the rate-PSNR curves with and without integrating BQE into G-PCC under the modified QP settings. It shows that BQE adapted well to these settings and achieved significant improvements.

\begin{figure*}
\centering
\includegraphics[width=6.2in]{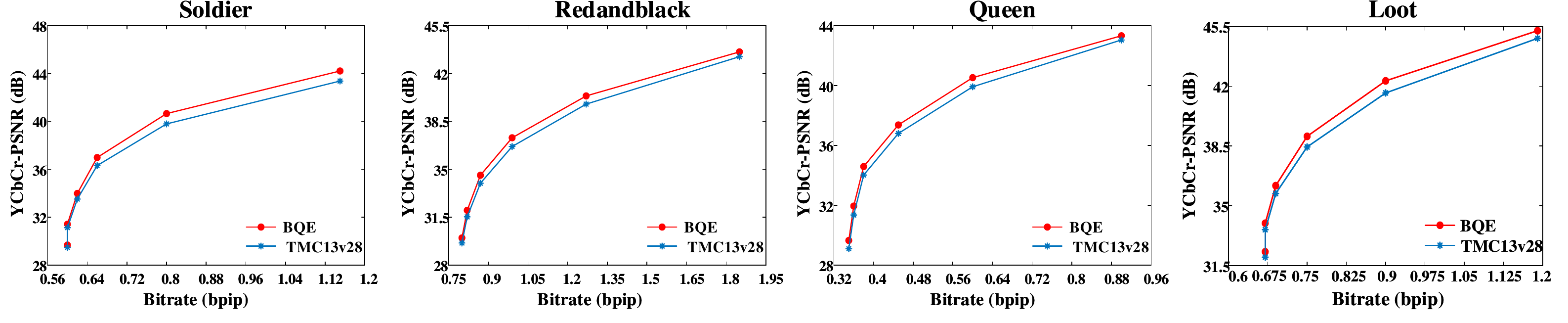}
\caption{Rate-PSNR curves before and after integrating BQE into G-PCC.}
\label{FIG4}
\end{figure*}

\begin{figure*}
\centering
\includegraphics[width=6.2in]{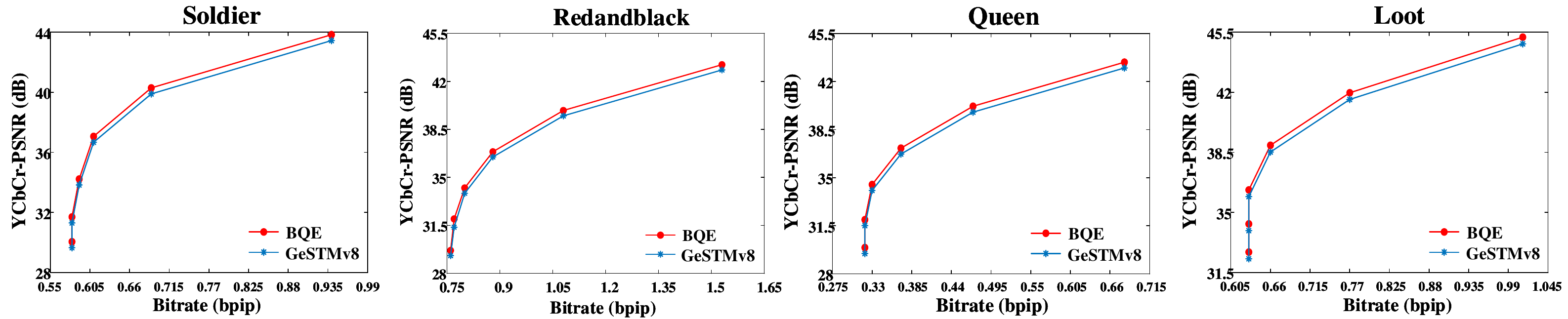}
\caption{Rate-PSNR curves before and after integrating BQE into GeSTMv8.}
\label{FIG5}
\end{figure*}

\begin{figure*}
\centering
\includegraphics[width=6.2in]{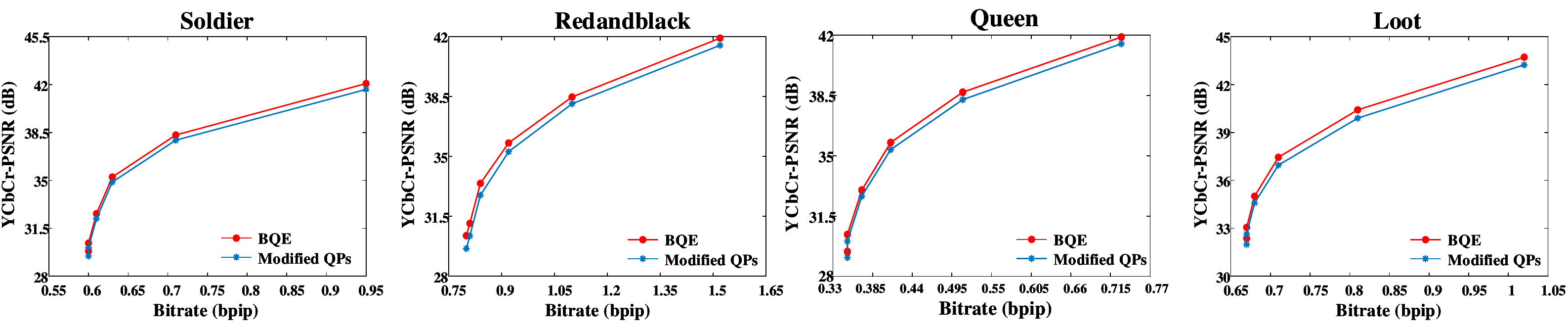}
\caption{Rate-PSNR curves before and after integrating BQE into G-PCC under modified QPs.}
\label{FIG6}
\end{figure*}
\begin{figure*}
\centering
\includegraphics[width=6.2in]{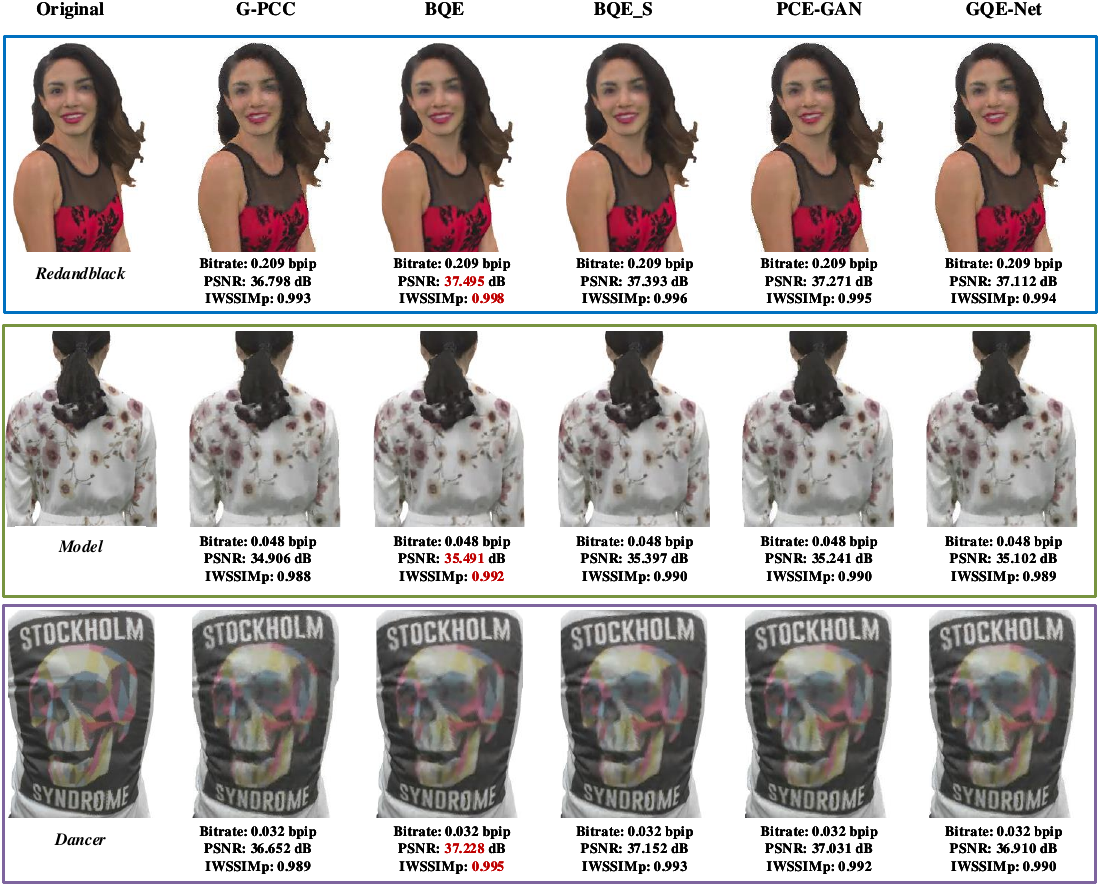}
\caption{Subjective quality comparison.}
\label{FIG7}
\end{figure*}

\subsection{Comparison with State-of-the-Art Methods}

To comprehensively evaluate the effectiveness of BQE, we compared it with two state-of-the-art learning-based point cloud quality enhancement methods: GQE-Net \cite{a5} and PCE-GAN \cite{a6}. Since these baselines operate on single-frame static point clouds, whereas BQE targets dynamic data, we constructed a fair static variant, BQE\_S, by removing the RMC and CA modules from BQE. In addition, because GQE-Net and PCE-GAN are non-blind methods and typically require one model per bitrate, we retrained each baseline as a single model using point clouds from all six bitrates for fairness. The average PSNRs and BD-rates of all tested sequences are reported in Table VII. The results show that BQE\_S achieved higher PSNR gains than GQE-Net and PCE-GAN, confirming the effectiveness of the proposed architecture in static scenarios. Moreover, by exploiting temporal information across frames, BQE further outperformed BQE\_S.
\begin{table*}[ht]
\centering
\caption{\(\Delta\)PSNR (dB) Comparison}
\label{tab:table7}  
\resizebox{15.5cm}{!}{
\begin{tabular}{lccc|ccc|ccc|ccc}
\toprule
\multirow{2}{*}{\textbf{Sequence}} 
& \multicolumn{3}{c}{\textbf{GQE-Net \cite{a5}}} 
& \multicolumn{3}{c}{\textbf{PCE-GAN \cite{a6}}} 
& \multicolumn{3}{c}{\textbf{BQE\_S}} 
& \multicolumn{3}{c}{\textbf{BQE}} \\
& \textbf{Y} & \textbf{Cb} & \textbf{Cr}
& \textbf{Y} & \textbf{Cb} & \textbf{Cr}
& \textbf{Y} & \textbf{Cb} & \textbf{Cr}
& \textbf{Y} & \textbf{Cb} & \textbf{Cr} \\
\hline
Dancer      & 0.220 & 0.119 & 0.380 & 0.329 & 0.151 & 0.403 & 0.401 & 0.108 & 0.214 & 0.461 & 0.153 & 0.254 \\
Model       & 0.213 & 0.158 & 0.415 & 0.343 & 0.231 & 0.476 & 0.397 & 0.136 & 0.268 & 0.443 & 0.221 & 0.422 \\
Redandblack & 0.243 & 0.262 & 0.285 & 0.378 & 0.388 & 0.210 & 0.379 & 0.304 & 0.417 & 0.512 & 0.435 & 0.508 \\
Soldier     & 0.290 & 0.403 & 0.460 & 0.410 & 0.338 & 0.422 & 0.413 & 0.420 & 0.379 & 0.581 & 0.485 & 0.474 \\
Loot        & 0.200 & 0.530 & 0.511 & 0.292 & 0.335 & 0.489 & 0.317 & 0.604 & 0.453 & 0.428 & 0.764 & 0.590 \\
Queen       & 0.168 & 0.398 & 0.352 & 0.376 & 0.474 & 0.505 & 0.437 & 0.354 & 0.484 & 0.513 & 0.532 & 0.615 \\
Phil        & 0.192 & 0.145 & 0.160 & 0.279 & 0.233 & 0.132 & 0.401 & 0.128 & 0.133 & 0.541 & 0.229 & 0.256 \\
Ricardo     & 0.224 & 0.306 & 0.349 & 0.257 & 0.202 & 0.434 & 0.528 & 0.329 & 0.424 & 0.634 & 0.376 & 0.520 \\
Sarah       & 0.255 & 0.305 & 0.287 & 0.344 & 0.308 & 0.302 & 0.576 & 0.446 & 0.362 & 0.705 & 0.433 & 0.437 \\
\hline
Average     & 0.223 & 0.292 & 0.356 & 0.334 & 0.296 & 0.375 & 0.427 & 0.314 & 0.348 & \textbf{0.535} & \textbf{0.403} & \textbf{0.453} \\
\bottomrule
\end{tabular}}
\end{table*}

\subsection{Subjective Quality Evaluation}

To evaluate subjective quality, we present visual comparisons on three sequences: \textit{Redandblack, Model}, and \textit{Dancer} (Fig. 7). We also used the full-reference point cloud quality metric (\(\text{IWSSIM}_{\text{P}}\)) \cite{IWSSIMP} to assess perceptual quality. Derived from IWSSIM \cite{IWSSIM}, this metric measures perceptual similarity in accordance with the human visual system. Fig. 7 shows that point clouds compressed by G-PCC suffered from severe compression artifacts. Compared with BQE\_S, PCE-GAN, and GQE-Net, BQE improved texture clarity. Notably, BQE achieved better results on the skin-fabric color transitions in \textit{Redandblack}, the hair boundaries in \textit{Model}, and the “STOCKHOLM” lettering in \textit{Dancer}.
\begin{table}[ht]
\centering
\caption{Computational Complexity Comparison}
\label{tab:table8}
\resizebox{8cm}{!}{
\begin{tabular}{lccc}
\toprule
\textbf{Method} & \textbf{Processing time (s)} & \textbf{FLOPs (G)} & \textbf{Parameters (M)} \\
\midrule
GQE-Net & 60.95 & 557.52 & 0.59 \\
PCE-GAN & 69.34 & 867.35 & 1.82 \\
BQE     & 22.06 & 303.31  & 0.71 \\
\bottomrule
\end{tabular}}
\end{table}

\subsection{Computational Complexity Analysis}

As summarized in Table VIII, BQE achieved the best overall efficiency: it required the least computation and had the shortest runtime, while maintaining a moderate model size. GQE-Net used fewer parameters but incurred substantially higher computational cost and runtime. PCE-GAN was the most resource-intensive, combining the largest model size with the highest computational cost.

\begin{table}[ht]
\centering
\caption{Ablation Study. \(R=0\) indicates that only the current frame is used as input to BQE, while \(R=1\) indicates that the two neighboring reference frames are also used.}
\label{tab:table9}
\resizebox{8.8cm}{!}{
\begin{tabular}{lccccccc}
\toprule
\textbf{Sequence} 
& \textbf{w/o TCCA} 
& \textbf{w/o PE} 
& \textbf{w/o NA} 
& \textbf{w/o QE} 
& \textbf{\(R=0\)} 
& \textbf{\(R=1\)} 
& \textbf{BQE} \\
\hline
Dancer      & 0.405 & 0.407 & 0.301 & 0.340 & 0.401 & 0.426 & 0.461 \\
Model       & 0.388 & 0.394 & 0.285 & 0.263 & 0.397 & 0.417 & 0.443 \\
Redandblack & 0.422 & 0.443 & 0.310 & 0.338 & 0.379 & 0.432 & 0.512 \\
Soldier     & 0.456 & 0.469 & 0.335 & 0.355 & 0.413 & 0.488 & 0.581 \\
Loot        & 0.310 & 0.359 & 0.270 & 0.248 & 0.317 & 0.334 & 0.428 \\
Queen       & 0.418 & 0.493 & 0.302 & 0.332 & 0.437 & 0.461 & 0.513 \\
Phil        & 0.439 & 0.429 & 0.318 & 0.342 & 0.401 & 0.458 & 0.541 \\
Ricardo     & 0.475 & 0.612 & 0.435 & 0.453 & 0.528 & 0.622 & 0.634 \\
Sarah       & 0.462 & 0.654 & 0.371 & 0.367 & 0.576 & 0.679 & 0.705 \\
\hline
Average     & 0.419 & 0.473 & 0.325 & 0.338 & 0.427 & 0.480 & 0.535 \\
\bottomrule
\end{tabular}}
\end{table}
\subsection{Ablation Study}

To verify the effectiveness of the proposed modules in BQE, we compared the performance of BQE with the following configurations:

\noindent (i) BQE \textbf{w/o TCCA}, i.e., we used a basic MLP \cite{b29} to replace TCCA.

\noindent (ii) BQE \textbf{w/o PE}, i.e., the positional encoding was removed from the NA module.

\noindent (iii) BQE \textbf{w/o NA}, i.e., we used a basic MLP to replace the NA module.

\noindent (iv) BQE \textbf{w/o QE}, i.e., the QE module was removed from BQE, and we directly set \(p_H=1\) and \(p_L=p_M=0\)

\noindent (v) \(R=0\), i.e., only the current frame \({\hat{\bm{P}}}_t\) was used as the input to BQE. 

\noindent (vi) \(R=1\), i.e., the current frame \({\hat{\bm{P}}}_t\) and its two reference frames \({\hat{\bm{P}}}_{t-1}\),\({\hat{\bm{P}}}_{t+1}\) are used as the input to BQE.

Table IX presents the results of the ablation study. Compared with the full BQE model, which achieved an average PSNR gain of 0.535 dB, removing the NA or QE module led to the most severe performance degradation, reducing the gain to 0.325 dB and 0.338 dB, respectively. This shows that neighborhood-aware feature extraction and distortion-aware adaptive fusion were crucial for effective quality enhancement. Removing the TCCA module or positional encoding also caused noticeable performance drops (to 0.419 dB and 0.473 dB, respectively), indicating that temporal correlation-guided cross-attention helped exploit temporal dependencies, while explicit positional encoding in the NA module exploited local geometric relationships to guide feature aggregation and improve the modeling of local structures. Regarding the number of input frames, using only the current frame (\(R=0\)) yielded a lower average gain of 0.427 dB, while including the current frame and two reference frames (\(R=1\)) improved the gain to 0.480 dB. Further increasing the temporal neighborhood to \(R=2\), that is, using the current frame together with four reference frames, provided the best performance, demonstrating that BQE consistently benefited from richer temporal context.

\section{Conclusion}
We proposed BQE, the first blind quality enhancement model for compressed point clouds. By jointly exploiting temporal dependencies and feature similarities and differences across multiple distortion levels, BQE overcomes the limitations of non-blind methods that require separate models for each QP. The BQE model consists of a joint progressive feature extraction branch and an adaptive feature fusion branch. The joint progressive feature extraction branch effectively captures representations across multiple distortion levels, while the adaptive feature fusion branch predicts distortion-aware weighting distributions to guide adaptive feature fusion. Experimental results demonstrate that BQE significantly improves the quality of compressed dynamic point cloud attributes without prior knowledge of the distortion level. In future work, we will extend BQE to joint geometry–attribute enhancement.

 

\begin{thebibliography}{1}
\bibliographystyle{IEEEtran}

\bibitem{z1} A. L. Souto, R. L. De Queiroz, and C. Dorea, ``Motion-compensated predictive RAHT for dynamic point clouds,'' \textit{IEEE Trans. Image Process.}, vol. 32, pp. 2428-2437, 2023.

\bibitem{z2} A. Akhtar, Z. Li, and G. Van der Auwera, ``Inter-frame compression for dynamic point cloud geometry coding,'' \textit{IEEE Trans. Image Process.}, vol. 33, pp. 584-594, 2024.

\bibitem{z3} D. C. Garcia, T. A. Fonseca, R. U. Ferreira, and R. L. de Queiroz, ``Geometry coding for dynamic voxelized point clouds using octrees and multiple contexts,'' \textit{IEEE Trans. Image Process.}, vol. 29, pp. 313-322, 2020.

\bibitem{z4} L. Li, Z. Li, V. Zakharchenko, J. Chen, and H. Li, ``Advanced 3D motion prediction for video-based dynamic point cloud compression,'' \textit{IEEE Trans. Image Process.}, vol. 29, pp. 289-302, 2020.

\bibitem{z5} Y. Qian, J. Hou, S. Kwong, and Y. He, ``Deep magnification-flexible upsampling over 3D point clouds,'' \textit{IEEE Trans. Image Process.}, vol. 30, pp. 8354-8367, 2021.

\bibitem{z6} W. Gao, L. Xie, S. Fan, G. Li, S. Liu, and W. Gao, ``Deep learningbased point cloud compression: an in-depth survey and benchmark,'' \textit{IEEE Trans. Pattern Anal. Mach. Intell.}, early access, 2025, doi:10.1109/TPAMI.2025.3594355.

\bibitem{z7} W. Gao, H. Yuan, G. Li, Z. Li, and H. Yuan, ``Low complexity coding unit decision for video-based point cloud compression,'' \textit{IEEE Trans. Image Process.}, vol. 33, pp. 149-162, 2024.

\bibitem{z8} Y. Zeng, J. Hou, Q. Zhang, S. Ren, and W. Wang, ``Dynamic 3D point cloud sequences as 2D videos,'' \textit{IEEE Trans. Pattern Anal. Mach. Intell.}, vol. 46, no. 12, pp. 9371-9386, 2024.

\bibitem{z9} T. Guo, C. Chen, H. Yuan, X. Mao, R. Hamzaoui, and J. Hou, ``CS-Net: contribution-based sampling network for point cloud simplification,'' \textit{IEEE Trans. Vis. Comput. Graphics}, vol. 31, no. 10, pp. 9154-9165, 2025.

\bibitem{z10} X. Mao, H. Yuan, T. Guo, S. Jiang, R. Hamzaoui, and S. Kwong, ``SPAC: sampling-based progressive attribute compression for dense point clouds,'' \textit{IEEE Trans. Image Process.}, vol. 34, pp. 2939-2953, 2025.

\bibitem{z11} J. He, C. Li, S. Wang, and S. Kwong, ``Improving robustness of point cloud analysis through perturbation simulation and distortion-guided feature augmentation,'' \textit{IEEE Trans. Image Process.}, vol. 34, pp. 6683-6698, 2025.

\bibitem{z12} W. Zhao, W. Gao, D. Li, J. Wang, and G. Liu, ``LOD-PCAC: level-of-detail-based deep lossless point cloud attribute compression,'' \textit{IEEE Trans. Image Process.}, vol. 34, pp. 3918-3929, 2025.

\bibitem{z13} S. Schwarz et al., ``Emerging MPEG standards for point cloud compression,'' \textit{IEEE J. Emerg. Sel. Topics Circuits Syst.}, vol. 9, no. 1, pp. 133-148, 2019.

\bibitem{z14} G-PCC codec description, document ISO/IEC JTC1/SC29/WG11 N19331, Apr. 2020.

\bibitem{z15} Y. Wang, Y. Sun, Z. Liu, S. E. Sarma, M. M. Bronstein, and J. M. Solomon, ``Dynamic graph CNN for learning on point clouds,'' \textit{ACM Trans. Graph. }, vol. 38, no. 5, pp. 146:1–146:12, 2019.

\bibitem{z16} Z. Li, S. Liu, W. Gao, G. Li, and G. Li, ``S4R: rethinking point cloud sampling via guiding upsampling-aware perception,'' \textit{IEEE Trans. Multimedia}, vol. 27, pp. 6677-6689, 2025.

\bibitem{z17} M. Simonovsky and N. Komodakis, ``Dynamic edge-conditioned filters in convolutional neural networks on graphs,'' in \textit{IEEE/CVF Conf. Comput. Vis. Pattern Recognit. }, 2017, pp. 29-38.

\bibitem{z18} M. Wei et al., ``AGConv: adaptive graph convolution on 3D point clouds,'' \textit{IEEE Trans. Pattern Anal. Mach. Intell.}, vol. 45, no. 8, pp. 9374-9392, 2023.

\bibitem{z19} S. Gu, J. Hou, H. Zeng, H. Yuan, and K. -K. Ma, ``3D point cloud attribute compression using geometry-guided sparse representation,'' \textit{IEEE Trans. Image Process.}, vol. 29, pp. 796-808, 2020.

\bibitem{z20} A. Akhtar, Z. Li, G. V. d. Auwera, L. Li, and J. Chen, ``PU-Dense: sparse tensor-based point cloud geometry upsampling,'' \textit{IEEE Trans. Image Process.}, vol. 31, pp. 4133-4148, 2022.

\bibitem{z21} X. Mao, H. Yuan, X. Lu, R. Hamzaoui, and W. Gao, ``PCAC-GAN: a sparse-tensor-based generative adversarial network for 3D point cloud attribute compression,'' \textit{Computational Visual Media}, vol. 11, no. 5, pp. 939-951, 2025.

\bibitem{z22} J. Wang, R. Xue, J. Li, D. Ding, Y. Lin, and Z. Ma, ``A versatile point cloud compressor using universal multiscale conditional coding – Part II: attribute,'' \textit{IEEE Trans. Pattern Anal. Mach. Intell.}, vol. 47, no. 1, pp. 252-268, 2025. 

\bibitem{a1} L. Wang, J. Sun, H. Yuan, R. Hamzaoui, and X. Wang, ``Kalman filter-based prediction refinement and quality enhancement for geometrybased point cloud compression,'' in \textit{Proc. Int. Conf. Vis. Commun. Image Process. }, Dec. 2021, pp. 1–5.

\bibitem{a2} J. Xing, H. Yuan, C. Chen, and T. Guo, ``Wiener filter-based point cloud adaptive denoising for video-based point cloud compression,'' in \textit{Proc. 1st Int. Workshop Adv. Point Cloud Compress., Process. Anal.}, Oct. 2022, pp. 21–25.

\bibitem{a3} T. Guo, H. Yuan, R. Hamzaoui, X. Wang, and L. Wang, ``Dependence based coarse-to-fine approach for reducing distortion accumulation in GPCC attribute compression,'' \textit{IEEE Trans. Ind. Informat.}, vol. 20, no. 9, pp. 11393–11403, Sep. 2024.

\bibitem{a4} X. Sheng, L. Li, D. Liu, and Z. Xiong, ``Attribute artifacts removal for geometry-based point cloud compression,'' \textit{IEEE Trans. Image Process.}, vol. 31, pp. 3399–3413, 2022.

\bibitem{a5} J. Xing, H. Yuan, R. Hamzaoui, H. Liu, and J. Hou, ``GQE-Net: a graph-based quality enhancement network for point cloud color attribute,'' \textit{IEEE Trans. Image Process.}, vol. 32, pp. 6303–6317, 2023.

\bibitem{a6} T. Guo, H. Yuan, Q. Liu, H. Su, R. Hamzaoui, S. Kwong, ``PCE-GAN: a generative adversarial network for point cloud attribute quality enhancement based on optimal transport,'' \textit{IEEE Trans. Image Process.}, vol. 34, pp. 6138-6151, 2025.

\bibitem{a7} P. Liu, W. Gao, and X. Mu, ``Fast inter-frame motion prediction for compressed dynamic point cloud attribute enhancement,'' in \textit{Proc. AAAI Conf. Artif. Intell.}, vol. 38, 2024, pp. 3720–3728. 

\bibitem{a8} J. Zhang, T. Chen, D. Ding, and Z. Ma, ``G-PCC++: enhanced geometry-based point cloud compression,'' in \textit{Proc. ACM Int. Conf. Multimedia}, Oct. 2023, pp. 1352–1363.

\bibitem{a9} J. Zhang, J. Zhang, D. Ding, and Z. Ma, ``Learning to restore compressed point cloud attribute: a fully data-driven approach and a rules-unrolling-based optimization,'' \textit{IEEE Trans. Vis. Comput. Graphics}, vol. 31, no. 4, pp. 1985–1998, Apr. 2025.

\bibitem{a10} J. Zhang, J. Zhang, D. Ding, and Z. Ma, ``ARNet: attribute artifact reduction for G-PCC compressed point clouds,'' \textit{Comput. Vis. Media}, vol. 11, no. 2, pp. 327–342, Apr. 2025.

\bibitem{b1} Y. Kim et al., ``A pseudo-blind convolutional neural network for the reduction of compression artifacts,'' \textit{IEEE Trans. Circuits Syst. Video Technol.}, vol. 30, no. 4, pp. 1121–1135, Apr. 2020 

\bibitem{JPEG} G. K.Wallace, ``The JPEG still picture compression standard,'' \textit{IEEE Trans. Consum. Electron.}, vol. 38, no. 1, pp. xviii–xxxiv, Feb. 1992.

\bibitem{H264} T. Wiegand, G. J. Sullivan, G. Bjontegaard, and A. Luthra, ``Overview of the H. 264/AVC video coding standard,'' \textit{IEEE Trans. Circuits Syst. Video Technol.}, vol. 13, no. 7, pp. 560–576, Jul. 2003.

\bibitem{b2} Q. Xing, M. Xu, T. Li, and Z. Guan, ``Early exit or not: resource-efficient blind quality enhancement for compressed images,'' in \textit{Proc. Eur. Conf. Comput. Vis.}, 2020, pp. 275–292.

\bibitem{b3} J. Jiang, K. Zhang, and R. Timofte, ``Towards flexible blind JPEG artifacts removal,'' in \textit{Proc. IEEE/CVF Int. Conf. Comput. Vis.}, 2021, pp. 4977-4986.

\bibitem{b5} J. Li, X. Liu, Y. Gao, L. Zhuo, and J. Zhang, ``BARRN: a blind image compression artifact reduction network for industrial IoT systems," \textit{IEEE Trans. Ind. Informat.}, vol. 19, no. 9, pp. 9479-9490, Sept. 2023.

\bibitem{b4} Q. Ding, L. Shen, L. Yu, H. Yang, and M. Xu, ``Blind quality enhancement for compressed video,'' \textit{IEEE Trans. Multimedia}, vol. 26, pp. 5782-5794, 2024.

\bibitem{b6} B. Li et al., ``PromptCIR: blind compressed image restoration with prompt learning,'' in \textit{Proc. IEEE/CVF Conf. Comput. Vis. Pattern Recognit. Workshops}, 2024, pp. 6442-6452.

\bibitem{b7} R. Yang et al., ``NTIRE 2024 challenge on blind enhancement of compressed image: methods and results,'' in \textit{Proc. IEEE/CVF Conf. Comput. Vis. Pattern Recognit. Workshops}, 2024, pp. 6524-6535.

\bibitem{z23} A. Vaswani et al., ``Attention is all you need,'' in \textit{Proc. Adv. Neural Inf. Process. Syst.}, 2017, pp. 6000–6010.

\bibitem{z24} H. Samet, ``K-nearest neighbor finding using maxnearestdist,'' \textit{IEEE Trans. Pattern Anal. and Mach. Intell.}, vol. 30, no. 2, pp. 243-252,2008. 

\bibitem{b20} E. d’Eon, B. Harrison, T. Myers, and P. A. Chou, ``8i voxelized full bodies, version 2 – a voxelized point cloud dataset,'' \textit{ISO/IEC JTC1/SC29 Joint WG11/WG1 MPEG/JPEG}, Input document m40059/M74006, Jan. 2017.

\bibitem{b21} Y. Xu, Y. Lu, and Z. Wen, ``Owlii dynamic human mesh sequence dataset,'' \textit{ISO/IEC JTC1/SC29/WG11 MPEG}, Input document m41658, Oct. 2017.

\bibitem{b22} C. Loop, Q. Cai, S. O. Escolano, and P. A. Chou, ``Microsoft voxelized upper bodies - a voxelized point cloud dataset,'' \textit{ISO/IEC JTC1/SC29 Joint WG11/WG1 MPEG/JPEG}, Input document m38673/M72012, May 2016. 

\bibitem{b25} Enhanced G-PCC test model v28, document ISO/IEC Standard JTC1/SC29/WG7 MPEG W24449, Dec. 2024.

\bibitem{b24} Common test conditions for G-PCC, document ISO/IEC Standard JTC1/SC29/WG7 MPEG N0368, Jul. 2022. 

\bibitem{b23} MPEG 3DG, 2018. [Online]. Available: \url{https://content.mpeg.expert/data/MPEG-I/Part05-PointCloudCompression/DataSets_New} 

\bibitem{b26} D. P. Kingma, and J. L. Ba, ``Adam: a method for stochastic optimization,'' in \textit{Proc. Int. Conf. Learn. Represent.}, 2015, pp. 1–15.

\bibitem{b27} Calculation of average PSNR differences between RD-curves, document Standard VCEG-M33, Apr. 2001. 

\bibitem{b28} Test model for geometry-based solid point cloud-GeSTMv8.0, document ISO/IEC Standard JTC1/SC29/WG7 MPEG MDS24474, Dec. 2024.

\bibitem{IWSSIMP} Q. Liu, H. Su, Z. Duanmu, W. Liu, and Z. Wang, “Perceptual quality assessment of colored 3D point clouds," \textit{IEEE Trans. Vis. Comput. Graphics}, vol. 29, no. 8, pp. 3642-3655, Aug. 2023.

\bibitem{IWSSIM} Z. Wang, and Q. Li, “Information content weighting for perceptual image quality assessment," \textit{IEEE Trans. Image Process.}, vol. 20, no. 5, pp. 1185-1198, May 2011. 

\bibitem{b29} D. E. Rumelhart, G. E. Hinton, and R. J. Williams, ``Learning representations by back-propagating errors,'' \textit{Nature}, vol. 323, no. 6088, pp. 533–536, 1986.  

\end{thebibliography}

\begin{IEEEbiography}[{\includegraphics[width=1in,height=1.25in,clip,keepaspectratio]{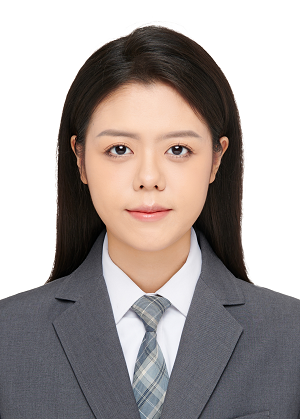}}]
 {Tian Guo}
 received the B.E. degree from the School of Information and Control Engineering, China University of Mining and Technology, Jiangsu, China, in 2021. She is currently pursuing the Ph.D. degree with the School of Control Science and Engineering, Shandong University, Shandong, China. Her research interests include point cloud compression and processing.
 \end{IEEEbiography}

 \begin{IEEEbiography}[{\includegraphics[width=1in,height=1.25in,clip,keepaspectratio]{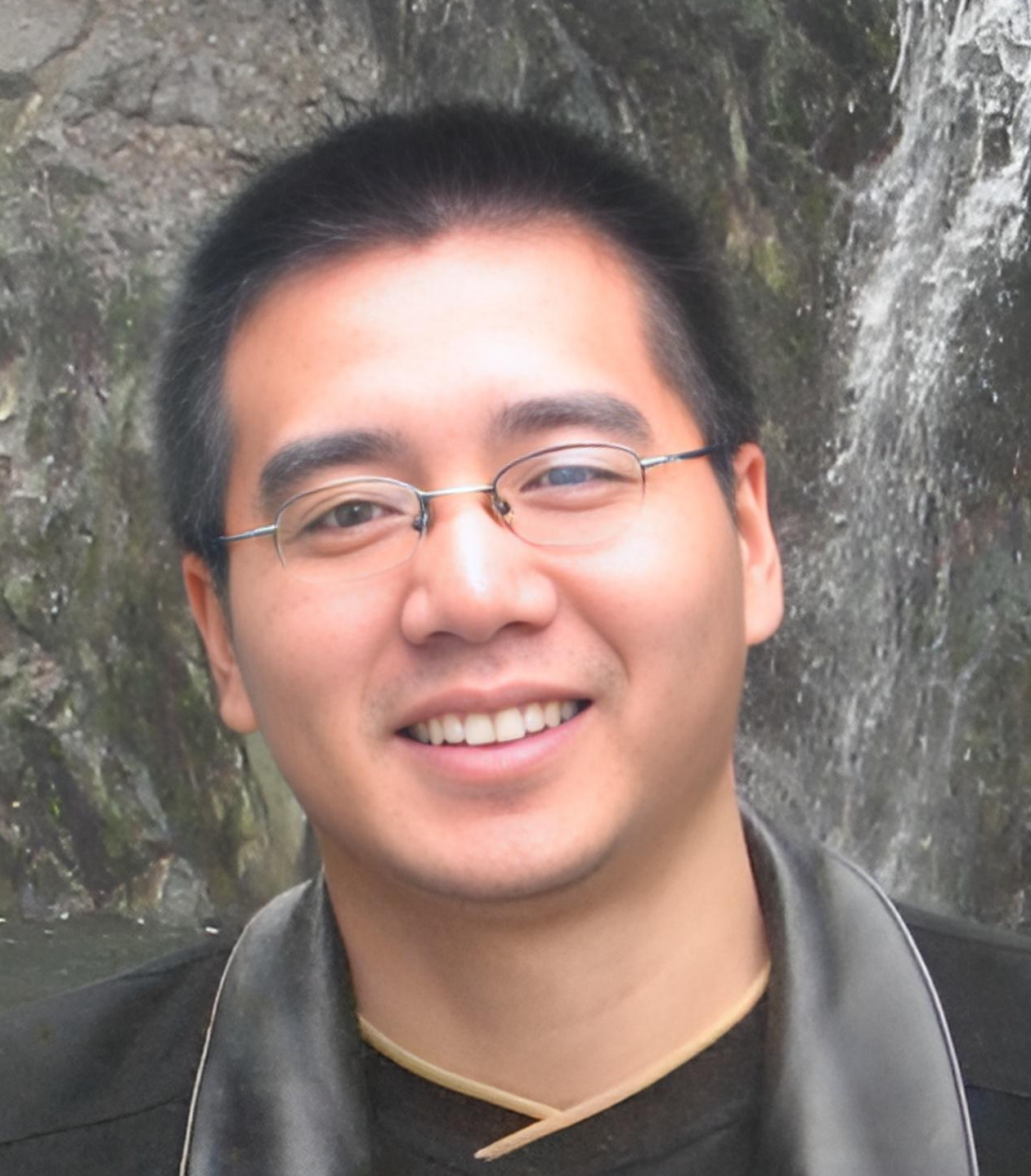}}]
 {Hui Yuan}
(Senior Member, IEEE) received the B.E. and Ph.D. degrees in telecommunication engineering from Xidian University, Xi’an, China, in 2006 and 2011, respectively. In April 2011, he joined Shandong University, Jinan, China, as a Lecturer (April 2011–December 2014), an Associate Professor (January 2015–August 2016), and a Professor (September 2016). From January 2013 to December 2014 and from November 2017 to February 2018, he was a Postdoctoral Fellow (Granted by the Hong Kong Scholar Project) and a Research Fellow, respectively, with the Department of Computer Science, City University of Hong Kong. From November 2020 to November 2021, he was a Marie Curie Fellow (Granted by the Marie Skłodowska-Curie Actions Individual Fellowship under Horizon2020 Europe) with the School of Engineering and Sustainable Development, De Montfort University, Leicester, U.K. From October 2021 to November 2021, he was also a Visiting Researcher (secondment of the Marie Skłodowska-Curie Individual Fellowships) with the Computer Vision and Graphics Group, Fraunhofer Heinrich-Hertz-Institut (HHI), Germany. His current research interests include 3D visual coding, processing, and communication. He is also serving as an Area Chair for IEEE ICME, an Associate Editor for \textit{IEEE Transactions on Image Processing}, \textit{IEEE Transactions on Consumer Electronics}, and \textit{IET Image Processing}.
 \end{IEEEbiography}

 \begin{IEEEbiography}[{\includegraphics[width=1in,height=1.25in,clip,keepaspectratio]{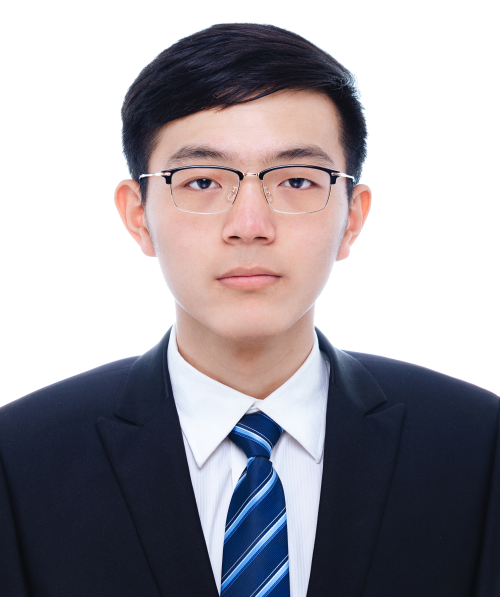}}]
 {Chang Sun}
received the B.S. and M.S. degrees in school of control science and engineering from Shandong University, Shandong, China, in 2019 and 2022. He is currently working toward the Ph.D degree in Shandong University. His research interests include point cloud compression and processing.
 \end{IEEEbiography}

 \begin{IEEEbiography}[{\includegraphics[width=1in,height=1.25in,clip,keepaspectratio]{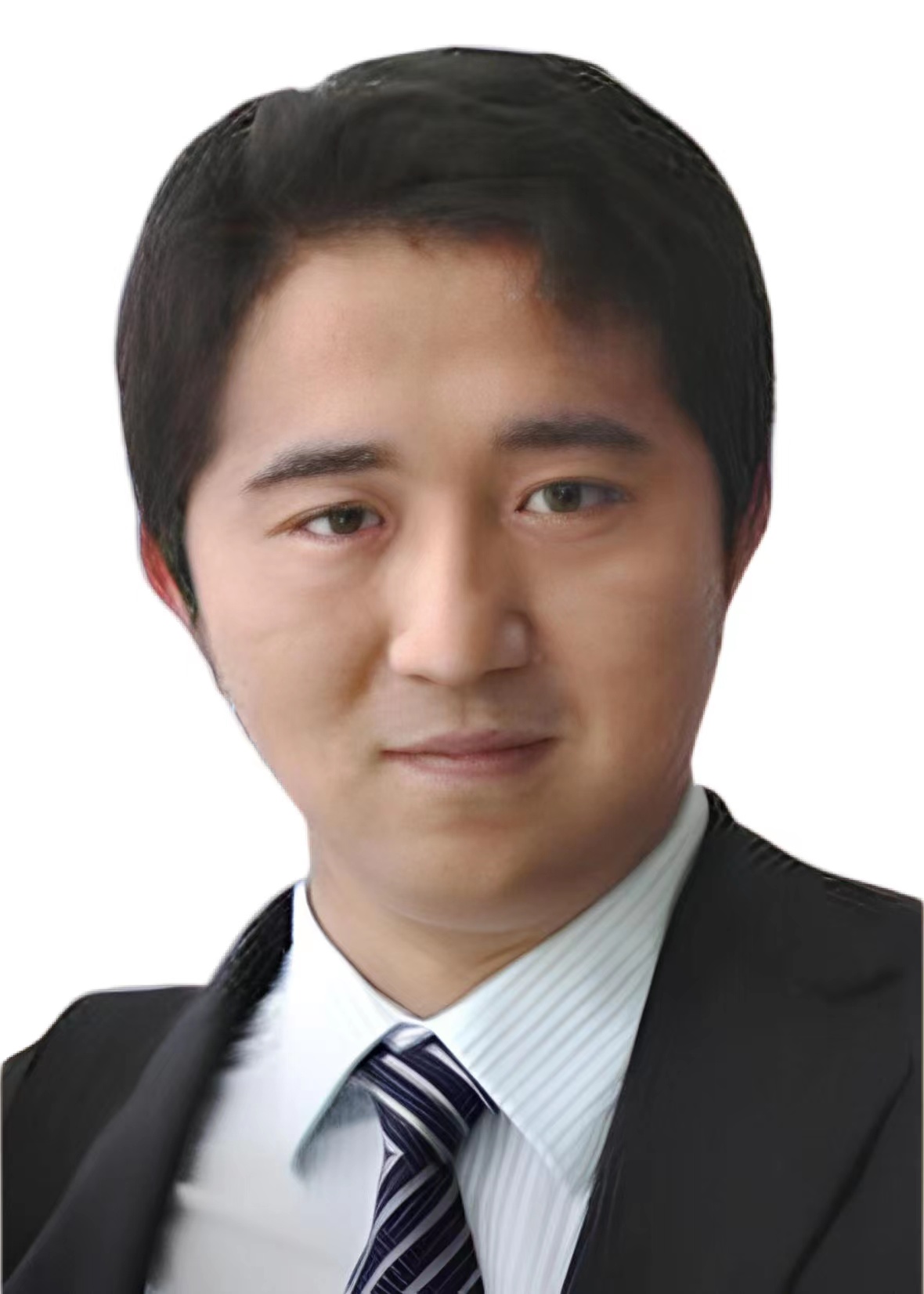}}]
{Wei Zhang} 
(Senior Member, IEEE) received the PhD degree in electronic engineering from the Chinese University of Hong Kong, in 2010. He is currently a professor with the School of Control Science and Engineering, Shandong University, China. He has published more than 120 papers in international journals and refereed conferences. His research interests include computer vision, image processing, pattern recognition, and robotics. He served as a program committee member and a reviewer for various international conferences and journals in image processing, computer vision, and robotics.
 \end{IEEEbiography}

 \begin{IEEEbiography}[{\includegraphics[width=1in,height=1.25in,clip,keepaspectratio]{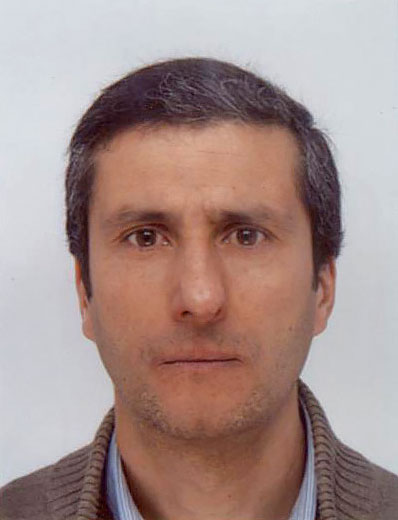}}]
 {Raouf Hamzaoui}
(Senior Member, IEEE) received the M.Sc. degree in mathematics from the University of Montreal, Canada, in 1993, the Dr.rer.nat. degree from the University of Freiburg, Germany, in 1997, and the Habilitation degree in computer science from the University of Konstanz, Germany, in 2004. He was an Assistant Professor with the Department of Computer Science, University of Leipzig, Germany, and the Department of Computer and Information Science, University of Konstanz. In September 2006, he joined De Montfort University, where he is currently a Professor in media technology. He was a member of the Editorial Board of the IEEE TRANSACTIONS ON MULTIMEDIA and IEEE TRANSACTIONS ON CIRCUITS AND SYSTEMS FOR VIDEO TECHNOLOGY. He has published more than 130 research papers in books, journals, and conferences. 
 \end{IEEEbiography}

 \begin{IEEEbiography}[{\includegraphics[width=1in,height=1.25in,clip,keepaspectratio]{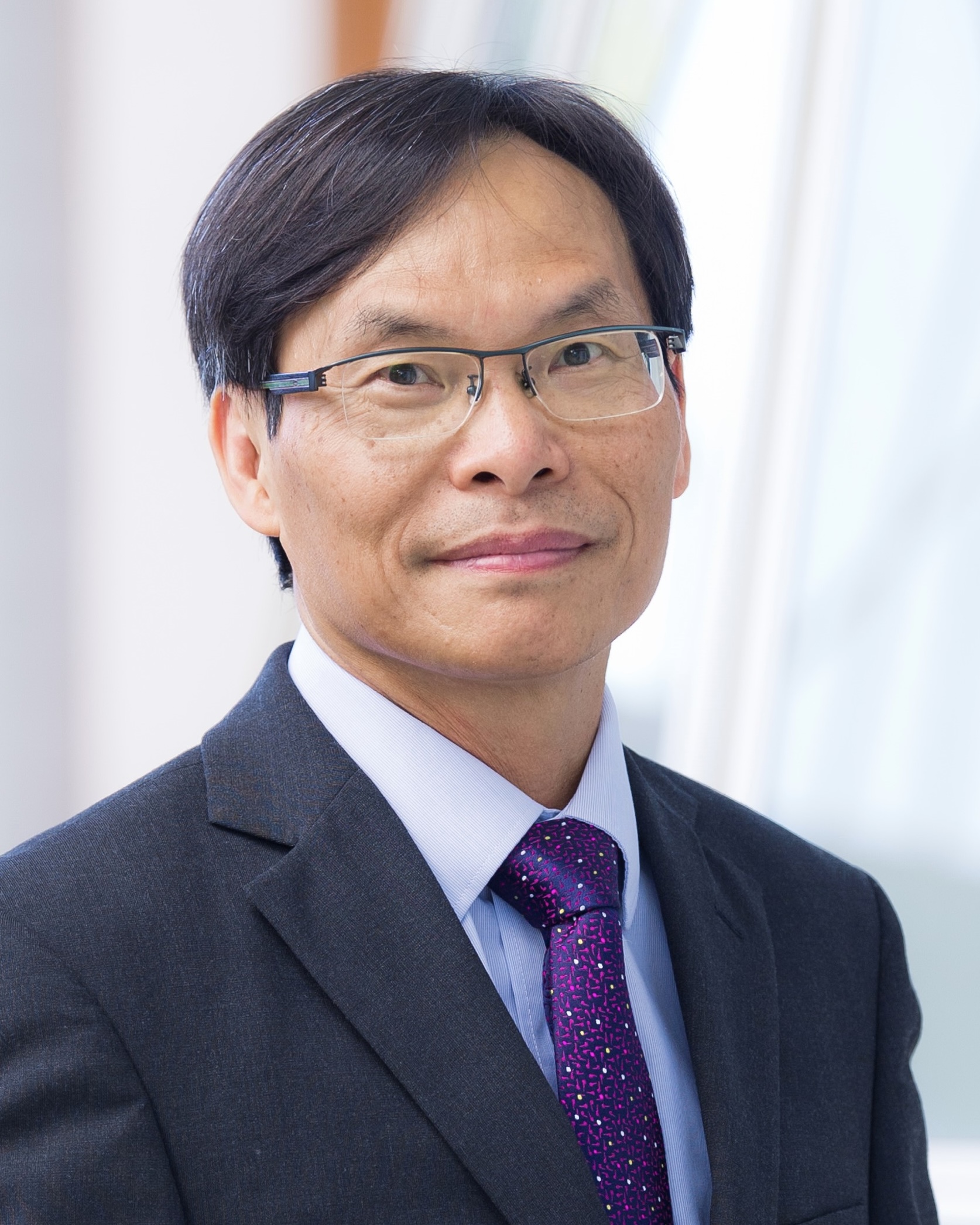}}]
 {Sam Kwong}
(Fellow, IEEE) is the Chair Professor of Computational Intelligence and concurrently serves as the Associate Vice-President (Strategic Research) at Lingnan University. He received the B.S. degree from the State University of New York at Buffalo in 1983, the M.S. degree in electrical engineering from the University of Waterloo, Canada, in 1985, and the Ph.D. degree from the University of Hagen, Germany, in 1996. From 1985 to 1987, he was a Diagnostic Engineer with Control Data Canada. He then joined Bell Northern Research Canada. Since 1990, he has been with City University of Hong Kong, where he served as a Lecturer in the Department of Electronic Engineering and later became a Chair Professor in the Department of Computer Science before moving to Lingnan University in 2023. His research interests include video/image coding, evolutionary algorithms, and artificial intelligence solutions. He is a Fellow of the IEEE, the Hong Kong Academy of Engineering Sciences (HKAES), and the National Academy of Inventors (NAI), USA. Dr. Kwong was honored as an IEEE Fellow in 2014 for contributions to optimization techniques in cybernetics and video coding and was named a Clarivate Highly Cited Researcher in 2022. He currently serves as an Associate Editor for the IEEE Transactions on Industrial Electronics and the IEEE Transactions on Industrial Informatics, among other prestigious IEEE journals. He has authored over 350 journal papers and 160 conference papers, achieving an h-index of 93 (Google Scholar). He served as President of the IEEE Systems, Man, and Cybernetics Society (SMCS) from 2021 to 2023.
 \end{IEEEbiography}

 




\vfill

\end{document}